\documentclass[aps,prb,final,reprint,twocolumn,floatfix,showpacs,superscriptaddress,notitlepage]{revtex4-1}
\usepackage{graphicx,amssymb,soul,color,amsmath,bm}
\usepackage{epstopdf,hyperref}
\usepackage{braket,dsfont}

\hypersetup{colorlinks=true,citecolor=blue,linkcolor=magenta}

\sethlcolor{yellow}
\allowdisplaybreaks

\begin{document}

\title{%
\textcolor{blue}{%
Nonequilibrium optical response of a one-dimensional Mott insulator
}%
}

\author{Juli\'an Rinc\'on}
\affiliation{School of Science, Engineering and Technology, Universidad del Rosario, Bogot\'a D.C. 111711, Colombia\looseness=-1}

\author{Adrian E.\ Feiguin}
\affiliation{Department of Physics, Northeastern University, Boston, Massachusetts 02115, USA}

\date{\today}

\begin{abstract}
We define, compute and analyze the nonequilibrium {\it differential} optical conductivity of the one-dimensional extended Hubbard model at half-filling after applying a pump pulse, using the time-dependent density matrix renormalization group method. The melting of the Mott insulator is accompanied by a suppression of the local magnetic moment and ensuing photogeneration of doublon-holon pairs. The differential optical conductivity reveals $(i)$ mid-gap states related to parity-forbidden optical states, and $(ii)$ strong renormalization and hybridization of the excitonic resonance and the absorption band, yielding a Fano resonance.
We offer evidence and interpret such a resonance as a signature of nonequilibrium optical excitations resembling excitonic strings, (bi)excitons, and unbound doublon-holon pairs, depending on the magnitude of the intersite Coulomb repulsion.
We discuss our results in the context of pump and probe spectroscopy experiments on organic Mott insulators.
\end{abstract}


\maketitle



\section{Introduction}
Coherent phenomena and its time-domain response in ultrafast timescales have been transformative in the understanding of both equilibrium and nonequilibrium electronic properties of quantum matter such as semiconductors, Mott insulators, and superconductors.~\cite{Orenstein_2012, Kampfrath_2013, Zhang_2014, Gandolfi_2017, Basov_2017, Wang_2018}
%
In Mott systems, decoherence of Mott-Hubbard excitons was detected in the one-dimensional (1D) organic salt ET-F$_2$TCNQ by performing pump-probe spectroscopy, where a THz oscillation in the reflectivity was interpreted as the quantum interference between exciton and unbound doublon-holon pairs.~\cite{Kampfrath_2013, Gandolfi_2017, Wall_2010} Exciton recombination in the same organic salt was also studied and the ensuing photodynamics was associated to the coherent evolution of holon-doublon pairs.~\cite{Mitrano14} Resonant ultrafast reflectivity measurements on the same molecular compound have shown the bleaching of the excitonic resonance and the photostabilization of biexcitons.~\cite{Miyamoto_2019} (See also Ref.~\onlinecite{Ono:2005aa, Okamoto07, Yamakawa_2017, Frenzel_2013, Lui:2014aa, Matsubara:2014aa}). Similarly, a photoinduced phase transition from charge density wave to Mott insulating states was observed in the 1D system [Pd(chxn)$_2$Br]Br$_2$ using femtosecond reflection spectroscopy.~\cite{Matsuzaki_2006} In this case, the nonequilibrium phase transition is caused by changes in the intersite Coulomb repulsion due to relaxation of Br atoms.

Understanding of such equilibrium and nonequilibrium phenomena in Mott insulators using theoretical methods has also been undertaken. Optical excitations in the 1D extended Hubbard model, whose ground state is a Mott insulator, can be due to the presence of excitons, biexcitons, excitonic strings, and charge-density-wave droplets.~\cite{Gebhard_1997a, *Gebhard_1997b, *Gebhard_1997c, Gallagher:1997aa, Kancharla:2001aa, Essler01, Jeck03}
Related theoretical efforts have explored the real-time dynamics of artificially created doublon-holon pairs,~\cite{Al-Hassanieh:2008aa} the ultrafast dynamics of recombination of doublon-holon pairs and excitons,~\cite{Lenarcic:2013aa, *Lenarcic:2015aa} the linear absorption using effective models~\cite{Ohmura:2019aa} in Mott insulators, the effect of such optical excitations on Coulombic screening,~\cite{Golez:2015aa} and the real-time dynamics of photoexcited electronic instabilities~\cite{Rincon:2014aa} and Hund excitons.~\cite{Rincon:2018aa}

Time-resolved analysis of optical excitations has led to advances on both $(i)$ the understanding of nonequilibrium physics and $(ii)$ the development of methods to describe effects of correlations in the time domain.
Generalizations of the optical conductivity, which is related to the reflectivity measured in spectroscopy experiments, have been introduced in the context of dynamical mean-field theory.~\cite{Eckstein:2008aa}
Other proposals aiming at mimicking the experimental setup of pump and probe experiments, resort to exact diagonalization to obtain the time-resolved optical conductivity.~\cite{Shao:2016aa} We will show below that this method actually computes the so-called differential optical conductivity~(\ref{eq:sigma}). This approach has been used to study the photogeneration of mid-gap states~\cite{Lu:2015aa} and quench-induced superconductivity~\cite{Paeckel:2020aa} in the 1D extended Hubbard model, and the relaxation dynamics via optical spectroscopy in the 1D Holstein model.~\cite{Kogoj:2016aa} 

In this work, we compute and analyze the time-domain optical response of a 1D correlated electron system described by the extended Hubbard model~(\ref{eq:H}) undergoing a pump pulse. Using time-dependent matrix product state (MPS) techniques,~\cite{dmrg1, dmrg2, dmrg3, Daley_2004, White04, Uli_2011, Paeckel2019} we register the time evolution of the ground state upon interaction with a light pulse. We determine the differential optical conductivity following Ref.~\onlinecite{Shao:2016aa} additionally taking into account the loss of significance when using tensor-network methods.

We notice a suppression of the local magnetic moment of the initial Mott insulating state indicating the photoexcitation of `hot' doublon-holon pairs which form both bound and deconfined optical excitations.
The main features of the differential optical conductivity are: $(i)$ the appearance of mid-gap states related to parity-forbidden states and $(ii)$ the strong renormalization and hybridization of the excitonic resonance and the absorption band.
We distinguish, depending on the value of the intersite Coulomb repulsion, nonequilibrium optical excitations resembling (bi)excitons, excitonic strings, and unbound doublon-holon pairs.
We also discuss the connections and differences between our approach and those used previously in the literature, as well as details on the numerical precision and use with tensor-network methods.

\section{Hamiltonian model}
We are interested in studying photoinduced optical excitations in a nominal one-dimensional Mott insulator after excitation with a pump light pulse. For this purpose, we consider the one-dimensional half-filled extended Hubbard model. The Hamiltonian reads
\begin{equation}
\begin{aligned}
H = &-\frac{W}{4}\sum\nolimits_{j, s} \left( c_{j s}^\dagger c_{j+1 s} + \textrm{H.c.} \right) \\
&+ U \sum\nolimits_j \left( n_{j\uparrow} - \frac{1}{2} \right) \left( n_{j\downarrow} - \frac{1}{2} \right) \\
&+ V \sum\nolimits_j \left( n_j - 1 \right) \left( n_{j+1} - 1 \right).\\
\end{aligned}
\label{eq:H}
\end{equation}
$W$, $U$, and $V$ are the bandwidth, local Hubbard repulsion, and nearest-neighbor repulsion, respectively. $c_{j s}^\dagger(c_{j s})$ creates (annihilates) an electron at lattice site $j$ with spin projection $s \in \{\uparrow, \downarrow\}$ and obey canonical commutation relations. $n_{js} = c_{j s}^\dagger c_{j s}$ and $n_j = n_{j\uparrow} + n_{j\downarrow}$ are number operators. Hamiltonian~(\ref{eq:H}) exhibits nontrivial optical excitations in equilibrium.~\cite{Gebhard_1997a, *Gebhard_1997b, *Gebhard_1997c, Gallagher:1997aa, Jeckelmann:2000aa, Kancharla:2001aa, Essler01, Jeck03}

The pump and probe light pulses that excite and probe the electron system~(\ref{eq:H}) are modeled as Gaussian envelopes with definite angular frequency $\omega$, width $\tau$, and amplitude $A$. We will henceforth use subscripts ``pmp'' and ``prb'' on those parameters to distinguish pump and probe pulses, respectively. The form of the pulses is
\begin{equation}
A(t) = A\, e^{-(t - t_{\rm peak})^2 / 2 \tau^2} \cos[\omega (t - t_{\rm peak})],
\label{eq:A}
\end{equation}
at $t_{\rm peak} = t_{\rm pmp},\, t_{\rm prb}$ the pulse has maximum strength.

The coupling between matter (electron system) and light (pulses) is performed in the velocity gauge using the transformation $c_{j s}^\dagger c_{j+1 s} \to e^{i A(t)} c_{j s}^\dagger c_{j+1 s}$, where $A(t)$ is the vector potential associated to the electric field of the light pulse $E(t) = - \partial_t A(t)$. This gauge transformation is also known as the Peierls substitution. With this coupling we define the electric current operator as
\begin{equation}
J(t) = \frac{\delta H}{\delta A} = -i \frac{W}{4} \sum\nolimits_{j, s} \left( e^{i A(t)} c_{j s}^\dagger c_{j+1 s} - \textrm{H.c.} \right).
\label{eq:J}
\end{equation}
%

\section{Differential optical conductivity\label{sec:diff_sigma}}
While the definition of the optical conductivity is well established in equilibrium, a single standard nonequilibrium definition is not available.~\cite{Lenarcic2014, Shao:2016aa, Eckstein2013, Rossini2014} In this section, we provide a unifying scheme that will allow us to define the {\it differential} optical conductivity, using the functional derivative from calculus of variations.~\cite{Ewing2016} Then we rederive a useful formula to calculate it numerically.~\cite{Shao:2016aa}

We consider a system far from equilibrium described by a many-body wave function $|\Psi(t)\rangle$ at time $t$. How the system was brought into this state is irrelevant in the following discussion, and could be due to a quench, a bias, or a pump pulse, for instance. In order to compute the linear response to a perturbation, we apply a weak electric pulse. The response of the system will be a function of the probing time and, presumably, shape and duration of the pulse.~\cite{Shao:2016aa} Analogously, the expectation value of the current~(\ref{eq:J}) will be a function of $t$ and $E$ so $J(t, E, t_{\rm prb}) = \langle\Psi(t)| J(t) |\Psi(t)\rangle$, where $t_{\rm prb}$ is the time at which the probe field is turned on. The corresponding Fourier transforms are (see Appendix~\ref{sec:fft} for details)
\begin{align}
E(\omega, t_{\rm prb}) &= \int_{t_{\rm prb}}^{\infty} dt\, e^{i\omega t} E(t, t_{\rm prb}),\\
J(\omega, E, t_{\rm prb}) &= \int_{t_{\rm prb}}^{\infty} dt\, e^{i\omega t} J(t, E, t_{\rm prb}).
\end{align}
%


%
%
Arriving to the definition of the differential optical conductivity requires the definition of the integrated current functional $\mathcal J$ as
\begin{equation}
\mathcal J(E) := \int_{0}^{\infty} d\omega \, J(\omega, E, t_{\rm prb}).
\label{eq:func}
\end{equation}
To calculate the functional derivative of $\mathcal J(E)$ with respect to $E$,~\cite{Ewing2016} we consider the variation $\delta E := \epsilon E_{{\rm prb}}$:
\begin{align*}
\int_{0}^{\infty} d\omega \, \frac{\delta \mathcal J}{\delta E} E_{\rm prb} &= \left[ \frac{d}{d\epsilon} \int_{0}^{\infty} d\omega \, J(\omega, E+\epsilon E_{\rm prb}, t_{\rm prb}) \right]_{\epsilon = 0} \\
&= \int_{0}^{\infty} d\omega \, \frac{\partial J}{\partial E} E_{\rm prb}.
\end{align*}
Comparing the first and last terms we obtain for the functional derivative of $\mathcal J(E)$
\begin{equation}
\frac{\delta \mathcal J}{\delta E} = \frac{\partial J}{\partial E}.
\end{equation}
Note that we are \emph{not} extremizing the functional $\mathcal J(E)$ and, therefore, we do \emph{not} demand that $\delta \mathcal J/\delta E = 0$.

On the other hand, we introduce the differential optical conductivity $\sigma(\omega, t_{\rm prb})$ by requiring that the functional differential of $\mathcal J(E)$ with respect to a variation of the electric field $E$ yields:
\begin{equation}
\delta \mathcal J = \int_{0}^{\infty} d\omega \, \sigma(\omega, t_{\rm prb}) E_{\rm prb}. 
\label{eq:diff}
\end{equation}
%

From these equations we obtain the explicit form of the \emph{differential optical conductivity}
\begin{equation}
\boxed{
	\sigma(\omega, t_{\rm prb}) = \frac{\delta \mathcal J}{\delta E} = \frac{\partial J}{\partial E}
}.
\label{eq:sigma}
\end{equation}
This result agrees with Ohm's law in the frequency domain $J(\omega) = \sigma(\omega) E(\omega)$ when there is a linear relation between the current and an electric field; $\sigma(\omega)$ is the optical conductivity. This is the reason why we have dubbed the equation above as the differential optical conductivity.
%
\footnote{Note that if $J = J(\omega, E, E_\omega, t_{\rm prb})$ is also a function of $E_\omega = \partial_\omega E$, variations of $E_\omega$ generalize~(\ref{eq:sigma}) to
$$
\sigma(\omega, t_{\rm prb}) 
 = \frac{\delta \mathcal J}{\delta E} 
= \frac{\partial J}{\partial E} - \frac{d}{d\omega} \frac{\partial J}{\partial E_\omega}.
$$

The consequences of this extra term will be explored in future work.}
The relation to linear response theory is discussed in Appendix~{\ref{sec:linear}}.

Let us now focus on the numerical calculation of $\sigma(\omega, \Delta t)$ in the context of a pump-probe experiment, where there are pump ($E_{\rm pmp}$) and probe ($E_{\rm prb}$) electric fields acting on the system. It is convenient to introduce a time delay $\Delta t = t_{\rm prb}-t_{\rm pmp}$ and rewrite~(\ref{eq:sigma}) as
\begin{align}
\sigma(\omega, \Delta t) &= \lim_{\epsilon \to 0} \frac{J(\omega, F+\epsilon E_{\rm prb}, t_{\rm prb}) - J(\omega,F,t_{\rm prb})}{\epsilon E_{\rm prb}} \nonumber \\
&=: \lim_{\epsilon \to 0} \sigma_\epsilon(\omega, \Delta t),
\label{eq:sigmae}
\end{align}
where we have introduced the numerically useful finite quotient $\sigma_\epsilon(\omega, \Delta t)$ of $\sigma(\omega, \Delta t)$. In this expression, $F$ is a field that could be due to a bias, for instance, while $J(\omega,F,t_{\rm prb})$ refers to the nonequilibrium current due to $E_{\rm pmp}$ but without applying $E_{\rm prb}$. (In the following, we set $F=0$.) The definition of the differential optical conductivity in terms of $\sigma_\epsilon(\omega, \Delta t)$ will allow us to numerically compute it by systematically varying $\epsilon$.

The optical conductivity at equilibrium $\sigma^{\rm eq}(\omega)$ can be obtained by setting $E_{\rm pmp} = 0$ such that
\begin{align}
\sigma^{\rm eq}(\omega) &= \lim_{\epsilon \to 0} \frac{J(\omega, \epsilon E_{\rm prb})}{\epsilon E_{\rm prb}} 
=: \lim_{\epsilon \to 0} \sigma^{\rm eq}_\epsilon(\omega).
\end{align}
In the equilibrium scenario there is no time delay variable and hence the optical conductivity only depends on frequency and on the probe field. Of course, in the limit $\epsilon \to 0$, $\sigma^{\rm eq}$ should not depend on the specific form of $E_{\rm prb}$. Lastly, note that the equilibrium optical conductivity can also be obtained from a pump and probe calculation as
\begin{equation}
\sigma^{\rm eq}(\omega) = \lim_{\Delta t\to -\infty} \sigma(\omega, \Delta t).
\end{equation}
This calculation is of course numerically more demanding, but it serves as a benchmark for the validity of the implementation of the computation of~(\ref{eq:sigma}) as $\sigma_\epsilon(\omega, \Delta t)$.

\section{Numerical method for \texorpdfstring{$\bm{\sigma(\omega, \Delta t)}$}{$\sigma(\omega, \Delta t)$}}
We now discuss the procedure to calculate the differential optical conductivity following Ref.~\onlinecite{Shao:2016aa}. The first step is to calculate $\sigma_\epsilon^{\rm eq}(\omega)$. In addition to providing information about optical excitations, it also establishes the range of values of $\epsilon$ such that $\epsilon E_{\rm prb}(\omega)$ can be considered a perturbation. That is, an electric field that does not perturb significantly $\sigma_\epsilon^{\rm eq}(\omega)$. The value of $\epsilon$ must so be chosen such that there is a balance between $\epsilon E_{\rm prb}(\omega)$ being an actual probe field while avoiding loss of significance as much as possible.

Once the appropriate value of $\epsilon$ has been set, we can proceed to the calculation of $\sigma(\omega, \Delta t)$. Let us suppose that we desire to compute $\sigma(\omega, \Delta t)$ in the time delay range $\Delta t \in [0, t_{\rm max}]$. We can discretize the interval in $N$ different bins with an effective time-delay step $t_{\rm max} / N$. Notice that we have the freedom to discretize such an interval into any number of points without impacting the accuracy of the calculation of~(\ref{eq:sigma}).

To calculate $\sigma(\omega, \Delta t)$, we use of the difference quotient $\sigma_\epsilon(\omega, \Delta t)$. For a given value of the pump and probe fields and the time delay $\Delta t$, we need to compute the currents (1) $J(\omega, E_{\rm pmp} + \epsilon E_{\rm prb})$ and (2) $J(\omega, E_{\rm pmp})$. This means that two different time-dependent density matrix renormalization group (DMRG) runs must be performed. However, one only needs one run to calculate $J(\omega, E_{\rm pmp})$ for several values of $E_{\rm prb}(\omega)$. Therefore, we need a linear number, $N + 1$, of different time-dependent runs to obtain $\sigma_\epsilon(\omega, \Delta t)$ in the discretized domain $[0, t_{\rm max}]$.

Let us estimate the time complexity of the whole algorithm. We assume a system of size $L$ with open boundary conditions, whose many-body wave function is represented by an MPS of bond dimension $\chi$. The discretized time domain consists of $N$ bins. Hence, the runtime to compute $\sigma_\epsilon(\omega, \Delta t)$ in the domain $\Delta t \in [0, t_{\rm max}]$ will require $O[(N + 2)L\chi^3 t_{\rm DMRG}]$ running time, where $t_{\rm DMRG}$ is the maximum time for a DMRG run. The additional simulation corresponds to the calculation of $\sigma_\epsilon^{\rm eq}(\omega)$. The DMRG simulation time must satisfy $t_{\rm DMRG} \gg t_{\rm max}$ such that Fourier transforms can reliably be performed, see Appendix~\ref{sec:fft}. Notice, additionally, that the time-delay step need not be uniformly distributed in the range $[0, t_{\rm max}]$. The time-dependent DMRG runs are independent from each other so the computation of $\sigma_\epsilon(\omega, \Delta t)$ can be embarrassingly parallelized.


\subsection*{Loss of significance in \texorpdfstring{$\bm{\sigma(\omega, \Delta t)}$}{$\sigma(\omega, \Delta t)$}}
Loss of significance, in {numerical differentiation}, occurs when subtracting two nearly equal floating-point numbers. This is always an important issue when approximating derivatives by finite quotients. In principle, we want $\epsilon$ as small as possible in numerical calculations. However, loss of significance introduces wild numerical fluctuations in the numerator of $\sigma_\epsilon(\omega, \Delta t)$. This is due to the loss of significant digits well beyond the accuracy of the numerical method used to obtain $J(\omega, E(\omega))$.

Whether $\sigma_\epsilon(\omega, \Delta t)$ is calculated using numerically exact methods, such as Lanczos, or using quasiexact methods, such as tensor networks, has a great impact on the results. In Lanczos, choosing $\epsilon$ at the limit of single precision ($\epsilon \sim 10^{-7}$) poses no problem since the error in these calculations is much smaller, close to machine precision, even when performing real-time evolution.

The situation is entirely different for tensor network methods. For instance, in time-dependent DMRG calculations, the truncation error is typically $\sim 10^{-8}-10^{-5}$ for sufficiently long times; expectation values have errors larger than the truncation error. The algorithm can then be conceived as a machine that outputs single-precision numbers so we can expect loss of significance when calculating the difference $J(E_{\rm pmp} + \epsilon E_{\rm prb}) - J(E_{\rm pmp})$ to quickly develop if $\epsilon$ is small enough.

Ideally, we would like to increase the bond dimension in order to improve the accuracy of DMRG's expectation values. Most of the time, however, this is not possible due to constraints in resources and time. We can work around having a finite bond dimension and avoid loss of significance by choosing `large' values of $\epsilon$. This has the effect of enhancing the response of the system to the probe field and concomitantly the accuracy of the expectation values computed. On the other hand, we need to choose a value of $\epsilon$ small enough such that the probe field is still a perturbation and does not alter the expectation values. The interplay between these two facts is the key to obtain reliable numerical estimates of $\sigma(\omega, \Delta t)$.

\section{Numerical results}
The results presented in this section are for the half-filled Hamiltonian~(\ref{eq:H}) with open boundary conditions. We employ time-dependent DMRG,~\cite{dmrg1, dmrg2, dmrg3, Daley_2004, White04, Uli_2011, Paeckel2019} which gives an MPS approximation to $|\Psi(t)\rangle$. Hamiltonian~(\ref{eq:H}) is time dependent via the vector potential~(\ref{eq:A}) and the evolution operator was approximated with a third-order Suzuki-Trotter expansion with time step $0.02$. We set $|\Psi(t = 0)\rangle$ to be the ground state of~(\ref{eq:H}) with $A = 0$ in~(\ref{eq:A}), which is obtained with ground-state DMRG. The calculations have been performed for system sizes up to $L = 48$ sites, maximum bond dimension $\chi = 600$, up to times $90$ and for parameters $W = 4$ (setting energy units), $U = 10$, varying $V = 1.5$, 3, and 4.5. The values of the probe field are $A_{\rm prb} = 0.05$, $\tau_{\rm prb} = 0.06$, $\omega_{\rm prb} = 10$, though similar results have been achieved with other values. For the pump field $A_{\rm pmp} = 0.3$, $\tau_{\rm pmp} = 0.5$, $t_{\rm pmp} = 2$. $\omega_{\rm pmp}$ was set to excite resonantly the lowest-energy equilibrium optical excitation according to the value of $V$ and $t_{\rm prb}$ was chosen according to the values of $\Delta t$ of interest. We observed that the truncation error always lies in the range $[10^{-8}, 10^{-4}]$, for the values of $V$, $\Delta t$, pump field, and largest values of $L$ studied. For more details we refer the reader to Refs.~\onlinecite{Al-Hassanieh:2008aa, Rincon:2014aa, Paeckel:2020aa}.

\subsection{Local magnetic moment}
\begin{figure}
\includegraphics*[width=.35\textwidth]{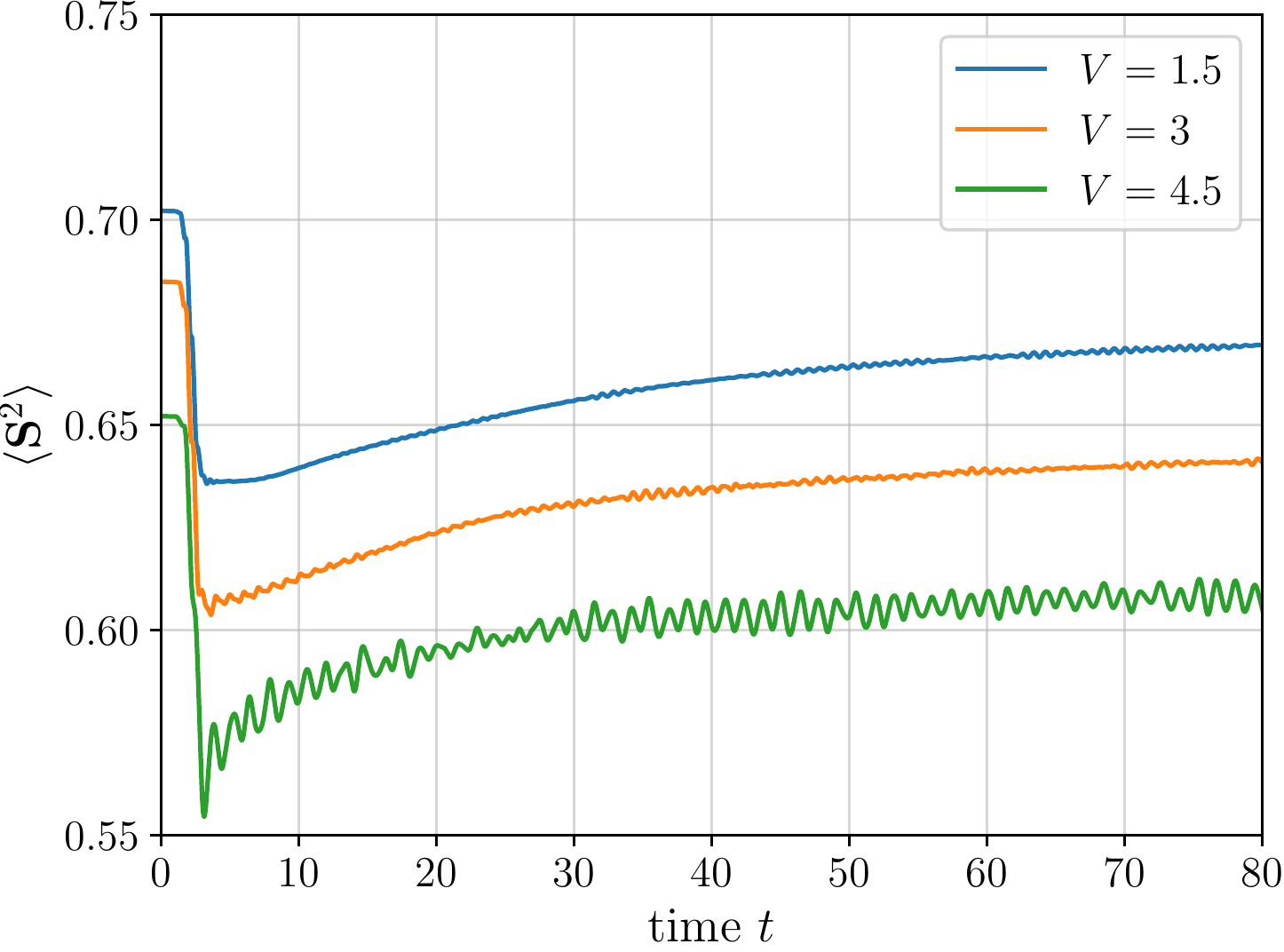}
\caption{Local magnetic moment $\langle \mathbf S^2 \rangle$ as a function of time $t$, for several values of the nearest-neighbor Coulomb interaction $V$. The span of the pump pulse lies in the range $t \in [0, 4]$ with maximum value at time $t_{\rm peak} = 2$. A partial melting of the magnetic moment is observed for all values of $V$.}
\label{fig:s2}
\end{figure}

Figure~\ref{fig:s2} shows the time dependence of the local magnetic moment $\langle \mathbf S^2 \rangle = \frac{1}{L}\sum_j \langle \Psi(t) | \mathbf S^2_j | \Psi(t) \rangle$ during and after the pump pulse is applied to the initial Mott insulating ground state. First, we notice that in equilibrium (at time $t = 0$), $\langle \mathbf S^2 \rangle$ diminishes with increasing $V$. Indeed, compared to the Hubbard-$U$ term in~(\ref{eq:H}), the term $V$ energetically favors double occupancy, hence lowering the effective magnetic moment. The relation between double occupancy $n_{j\uparrow}n_{j\downarrow}$ and $\mathbf S^2_j$ at site $j$ is
\begin{equation}
\mathbf S^2_j = \frac{3}{2} \left( \frac{1}{2} n_j - n_{j\uparrow}n_{j\downarrow} \right) \geqslant 0;
\end{equation}
therefore, an increase in the double occupancy implies a reduction of the local magnetic moment.

The quantity $\langle \mathbf S^2 \rangle$ indicates a partial melting of the magnetic moment associated to the equilibrium Mott insulating ground state, see Fig.~\ref{fig:s2}. The melting of $\langle \mathbf S^2 \rangle$ is accompanied by a reduction of antiferromagnetic correlations (not shown),~\cite{Matsueda:2005aa, Kohno:2010aa} and it is more pronounced when the pump pulse is being applied $t \in [0, 4]$. We observe an overall similar trend for different values of $V$, where the action of the resonant pump pulse reduces the magnetic moment by up to $7\,\%$ in the long-time regime. The period of the oscillations in $\langle \mathbf S^2 \rangle$, for different $V$, cannot be related to any energy scale. The reduction of the local magnetic moment implies a proliferation of doublons and holons. The transfer of energy from the pump pulse to the system makes high-energy states available during its time evolution. Such states typically contain a larger number of doublon-holon pairs compared to the ground state. Indeed, doublon-holon pairs are the excitations that drive the response observed in the time-resolved optical conductivity, as we will see below.

\begin{figure}
\includegraphics*[width=.4\textwidth]{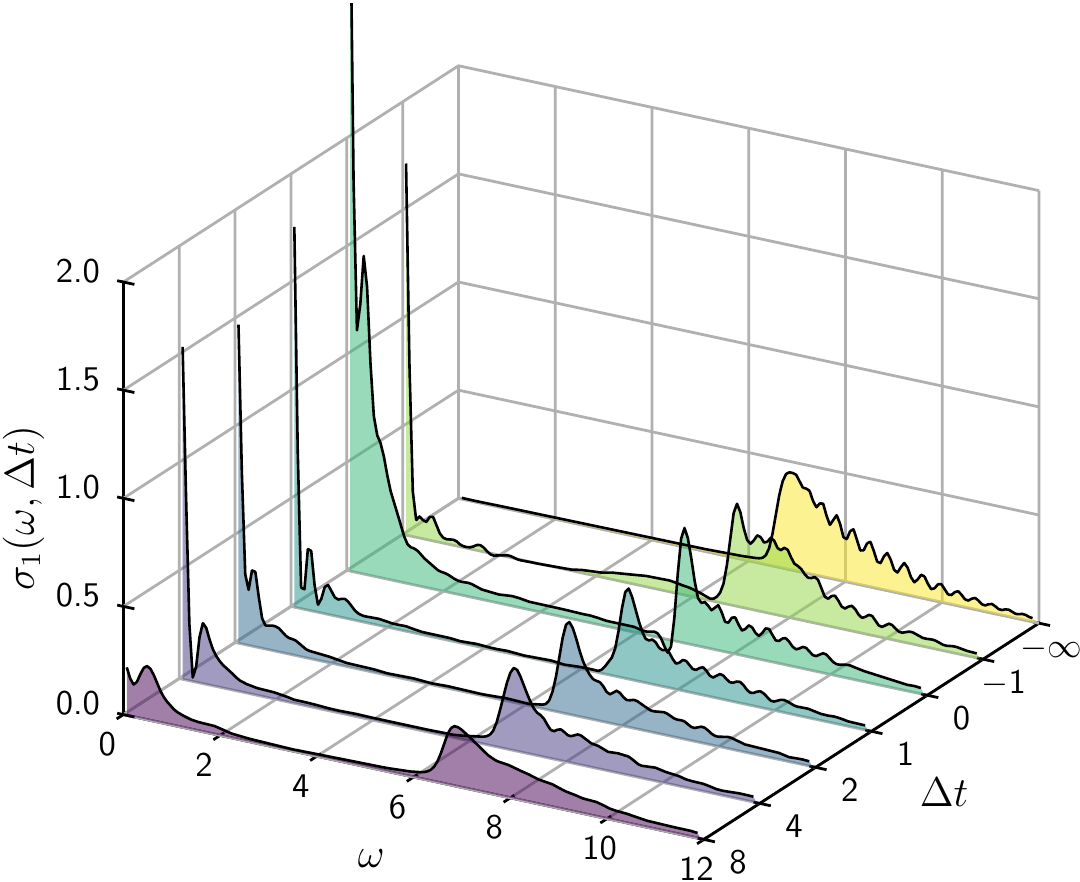}
\includegraphics*[width=.4\textwidth]{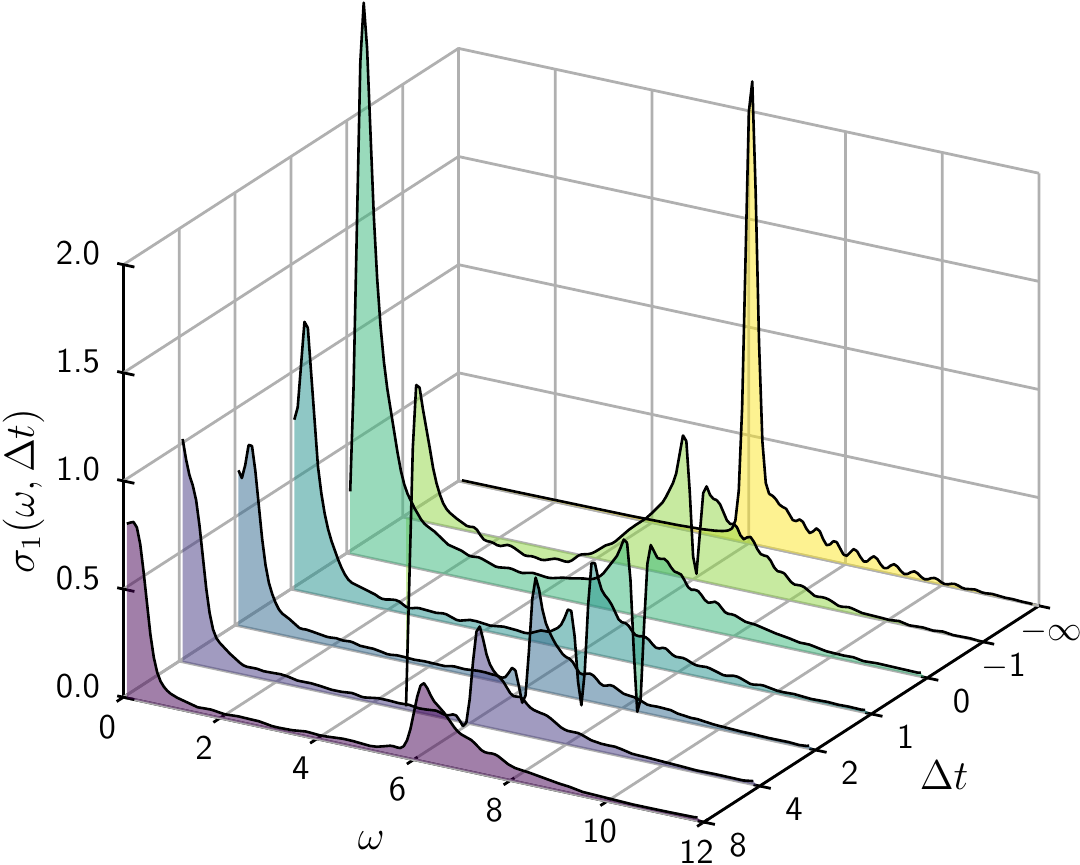}
\includegraphics*[width=.4\textwidth]{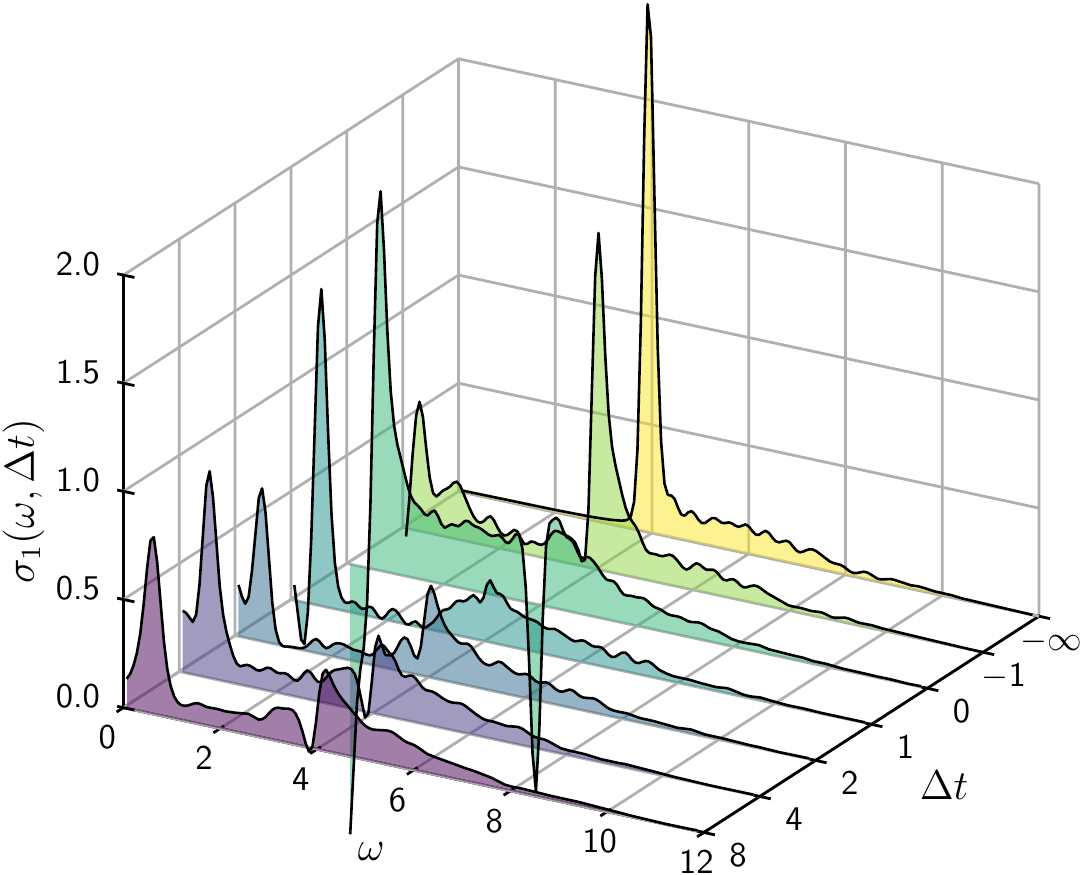}
\caption{Real part of the differential optical conductivity $\sigma_\epsilon(\omega, \Delta t)$ as a function of energy $\omega$ and pump-probe time delay $\Delta t$, for $V = 1.5$, 3, and 4.5 (from top to bottom). The time delay $\Delta t = -\infty$ corresponds to the equilibrium optical conductivity $\sigma_\epsilon^{\rm eq}(\omega)$. A convolution with a Gaussian window function was performed, see Appendix~\ref{sec:fft}.}
\label{fig:sigmat}
\end{figure}

\subsection{Differential optical conductivity}
The real part of the differential optical conductivity $\sigma_\epsilon(\omega, \Delta t)$ is shown in Fig.~\ref{fig:sigmat} for different values of the pump-probe time delay $\Delta t$. This quantity is calculated as the numerical version~(\ref{eq:sigmae}) of the differential optical conductivity~(\ref{eq:sigma}). The differential optical conductivity is defined as the variation of the current due to a pump pulse in the presence of a probe pulse. As discussed in Sec.~\ref{sec:diff_sigma} and other places,~\cite{Lu:2015aa, Shao:2016aa, Paeckel:2020aa} such definition of the optical conductivity allows to make a direct connection with pump and probe spectroscopy experiments.~\cite{Okamoto07, Wall_2010, Yamakawa_2017, Matsubara:2014aa}

The results for $\sigma_\epsilon(\omega, \Delta t)$ are plotted for three different values of the nearest-neighbor interaction, namely, $V = 1.5$, 3, and 4.5, in Fig.~\ref{fig:sigmat} (top to bottom). The equilibrium optical conductivity $\sigma_\epsilon^{\rm eq}(\omega)$ is marked as the time delay $\Delta t = -\infty$. For $V = 1.5$, a well-defined absorption band can be distinguished in the energy range $\omega \in [U-W, U+W]$ in agreement with previous work.~\cite{Gallagher:1997aa, Kancharla:2001aa, Essler01, Jeckelmann:2000aa, Jeck03} The spectral weight within such an interval is related to optical excitations composed of unbound doublon-holon pairs. For $V = 3$ and 4.5, we observe a well-defined peak with some spectral weight to the immediate right of it. The peak corresponds to bound optical excitations: (bi)excitons for $V = 3$ and excitonic strings for $V = 4.5$.~\cite{Gallagher:1997aa, Kancharla:2001aa, Essler01, Jeck03} The spectral weight spanning to the right of the excitonic peak corresponds to the doublon-holon absorption band; this would be the analog of the well-defined absorption band for $V = 1.5$. The optical gap is defined by the excitonic peak when $V \geqslant W/2$ and by the low-energy edge of the absorption band for $V < W/2$.~\cite{Kancharla:2001aa, Essler01, Jeck03} The results for the $\sigma_\epsilon^{\rm eq}(\omega)$ show that certainly the ground state corresponds to an insulator since there is no spectral weight at $\omega = 0$ for all values of $V$.

\begin{figure*}
\includegraphics*[width=.333\textwidth]{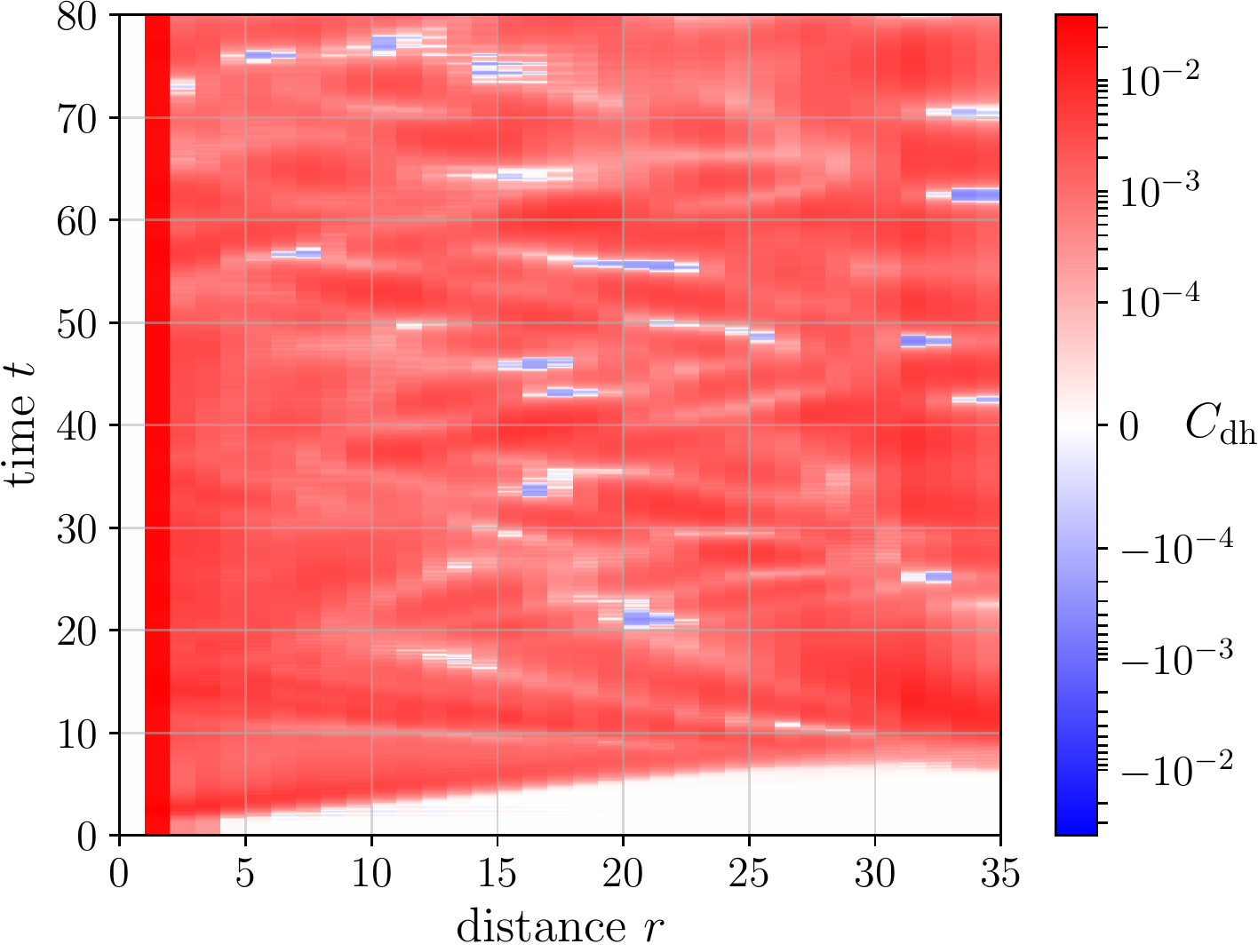}
\hspace{-.21cm}
\includegraphics*[width=.333\textwidth]{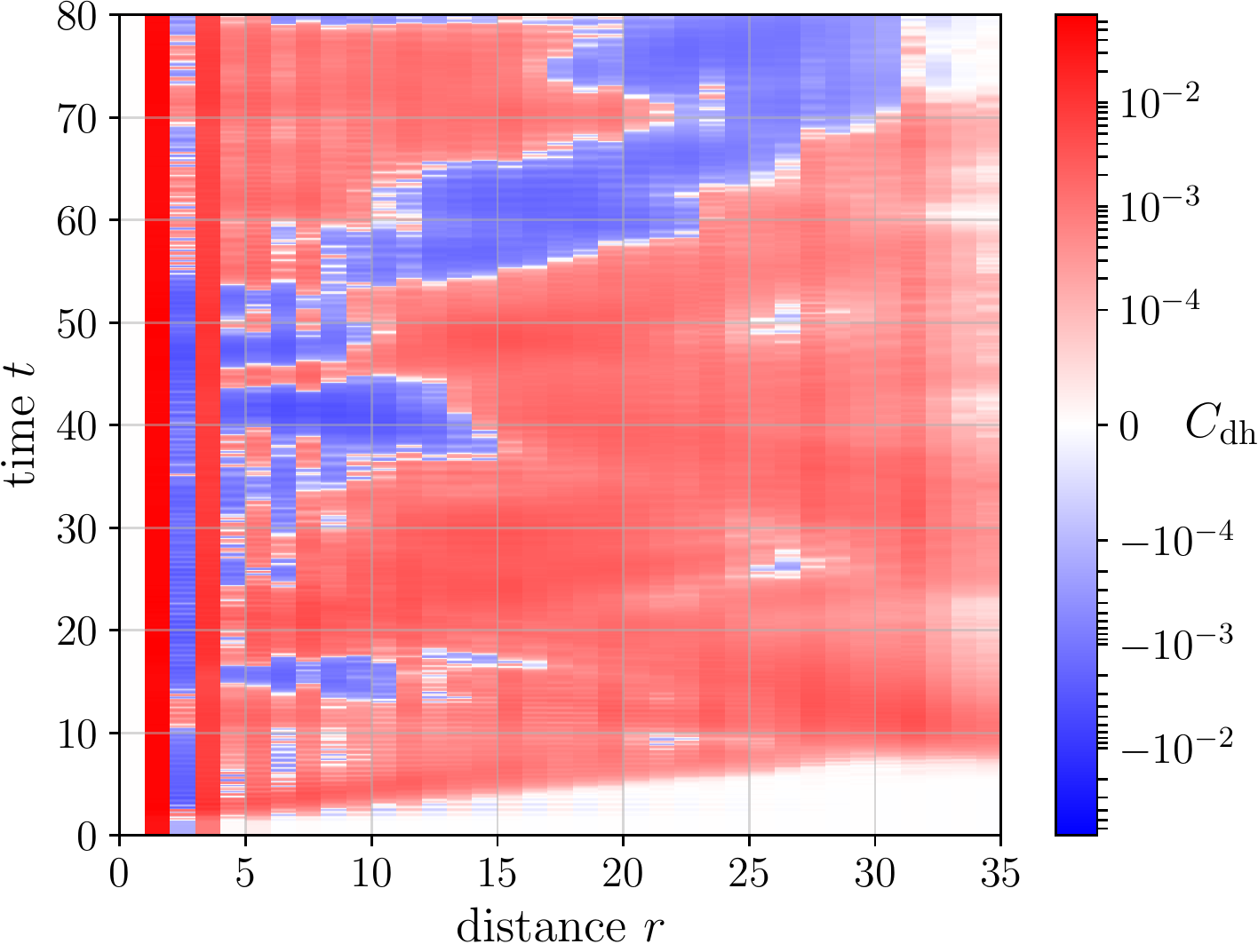}
\hspace{-.21cm}
\includegraphics*[width=.333\textwidth]{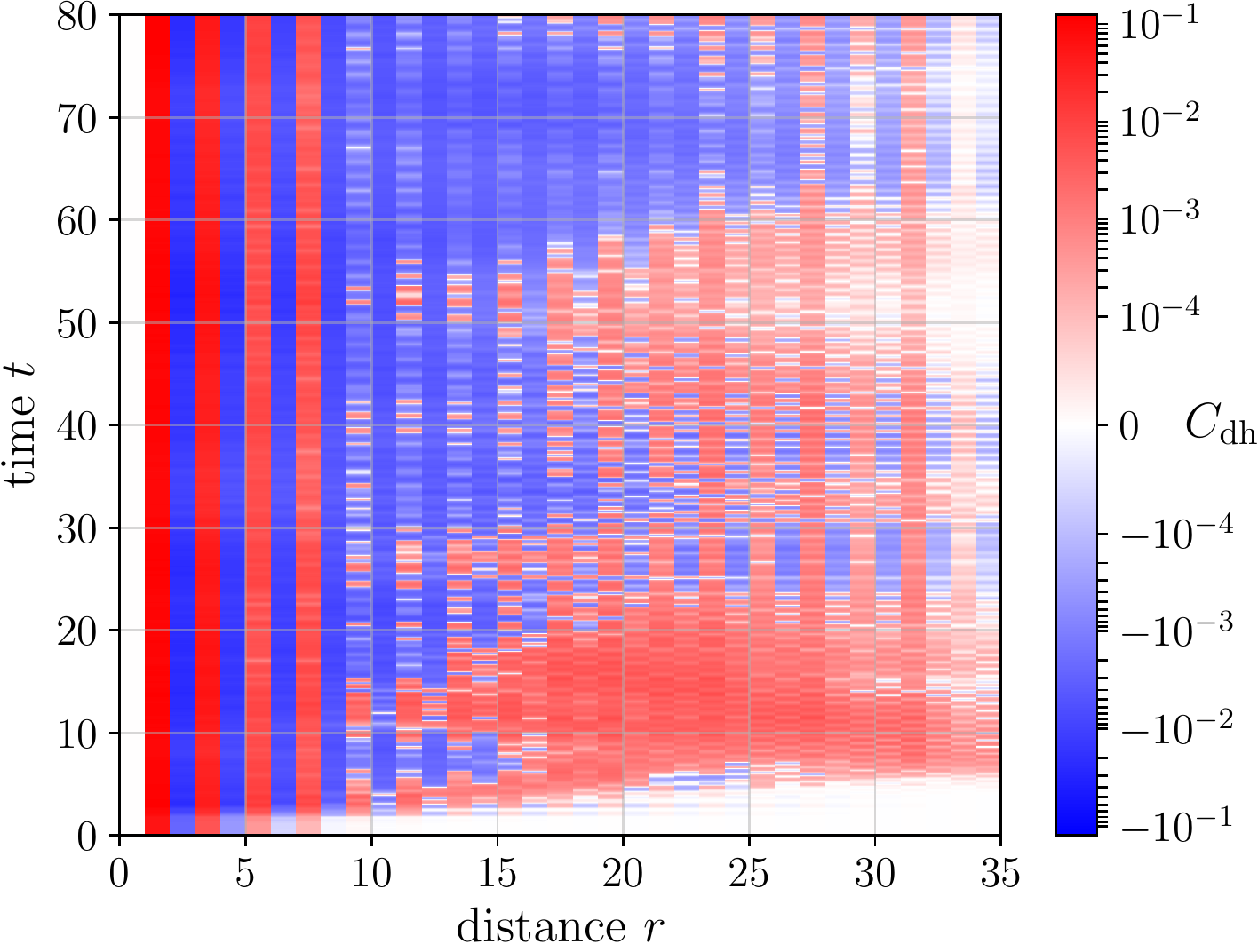}
\caption{Connected doublon-holon correlation $C_{\rm dh}(r, t)$ as a function of distance $r$ and time $t$ for the same values of $V$ shown in Fig.~\ref{fig:sigmat} (notice the logarithmic scale). From left to right: $V = 1.5$, 3, and 4.5. See text for further details.}
\label{fig:dhcorrt}
\end{figure*}

Let us now focus on the differential optical conductivity, shown in Fig.~\ref{fig:sigmat}. For $V < W/2$, for all the reported time delays $\Delta t$, we notice the rapid formation of an absorption band that somehow resembles that of equilibrium, albeit with a slight redistribution of its spectral weight. The sharp response of the electron system is a resonant effect of the pump pulse, which is tuned to the frequency of maximum spectral weight in the equilibrium absorption band, $\omega_{\rm pmp} \sim 7.25$. We will see below that this photoinduced absorption band is related to `hot' unbound doublon-holon quasiparticles (see Fig.~\ref{fig:dhcorrt}).

At energies below the optical gap and as $\Delta t$ increases, we observe the progressive formation of mid-gap states due to the transfer of optical spectral weight of the absorption band towards the low-$\omega$ region, see Fig.~\ref{fig:sigmat}. The signatures of photogenerated mid-gap states are reduced if the pump pulse is not resonant with the equilibrium absorption band (not shown). As discussed before,~\cite{Lu:2015aa, Shao:2016aa} the mid-gap states are related to optically dark states now accessible due to the coupling of the system with the pump pulse. Indeed, the wave function $|\Psi(t)\rangle$ allows to access optical states with forbidden parity; such dark states are manifested as mid-gap states.

We now turn to the differential optical conductivity for $V > W/2$. At small time delays $\Delta t \sim 0$, we observe a fast photoexcitation of the excitonic peak and a gradual photogeneration of mid-gap states. As in the case $V < W/2$, the fast response is due to the resonant tuning of the pump pulse to the excitonic peak. At large time delays $\Delta t \gg 1$, there is a robust mid-gap state and the accumulation of spectral weight around the excitonic resonance energy. Note, however, that this resonance now has a particular asymmetric shape (more pronounced for $V = 4.5$). At intermediate time delays $\Delta t \gtrsim 1$ we see a tendency towards a photometallic state, noticed as an accrual of spectral weight for $\omega \to 0^+$, that disappears at later time delays.

During the time evolution of the excitonic resonance, dips or negative spectral weight develop. Such features can be understood as the result of quantum interference between the excitonic resonance and the absorption band that lies right above it. A minimal effective model that captures the essential physics is the Fano Hamiltonian:~\cite{Fano:1961aa}
\begin{equation}
\tilde H = E_X |X\rangle\langle X| + \sum_k \varepsilon_k |k\rangle\langle k|
+ \sum_k \bigl( V_k(t) |X\rangle\langle k| + \mathrm{H.c.} \bigr),
\label{eq:fano}
\end{equation}
where $|X\rangle$ and $|k\rangle$ represent the excitonic state and the unbound doublon-holon quasiparticles that compose the absorption band, respectively. $E_X$ and $\varepsilon_k$ correspond to the energies of the exciton and the absorption band. $V_k$ is the hybridization between the excitonic resonance and the absorption band. Based on Fig.~\ref{fig:sigmat}, we observe that neither the excitonic resonance nor the absorption band change significantly their energy, we therefore attribute the time dependence to the hybridization coupling $V_k$.~\footnote{Further study of this model is currently underway.}

With Hamiltonian~(\ref{eq:fano}) we can unfold the features observed in $\sigma_\epsilon(\omega, \Delta t)$. Namely, the dip/negative spectral weight for small $\Delta t$ and the asymmetric absorption band at large $\Delta t$. Those two features are part of the same phenomenon: a light-induced Fano resonance. The asymmetric absorption profile seen around the excitonic energy for $\Delta t \gg 1$ is a characteristic signature of the Fano resonance.~\cite{Fano:1961aa} We will see below that the resulting asymmetric absorption band indeed corresponds to a nonequilibrium version of (bi)excitons and excitonic strings seen at equilibrium.~\cite{Essler01, Jeck03}


We notice that our results have direct connection with time-resolved THz spectroscopy experiments on the organic salt ET-F$_2$TCNQ, where quantum interference between excitons and absorption-band states have given rise the nonequilibrium optical excitations and photometallization.~\cite{Wall_2010, Mitrano14, Yamakawa_2017, Okamoto07, Miyamoto_2019}

\subsection{Doublon-holon correlations}
In order to shed light on the nature of the spectral peaks detected in $\sigma_\epsilon(\omega, \Delta t)$, we calculate the time-dependent connected doublon-holon correlation function at distance $r = m - j > 0$ and time $t$
\begin{equation}
C_{\rm dh}(r, t) = \langle d_j h_m + h_j d_m \rangle - \langle d_j \rangle \langle h_m \rangle - \langle h_j \rangle \langle d_m \rangle.
\label{eq:cdh}
\end{equation}
The expectation value is taken over $|\Psi(t)\rangle$, $d_j = n_{j\uparrow} n_{j\downarrow}$ and $h_j = (1 - n_{j\uparrow})(1 - n_{j\downarrow})$ are the double-occupancy and the hole number operators at lattice site $j$. The correlator $C_{\rm dh}(r, t)$ is a useful expectation value in the study of optical excitations generated by the current operator~(\ref{eq:J}).

Results for the connected doublon-holon correlation function are shown in Fig.~\ref{fig:dhcorrt} and varying $V$. The features shared by $C_{\rm dh}(r, t)$ for all values of $V$ shown are $(i)$ at short times, there is a ballistic build up of correlations from short to large distances $r$; and $(ii)$ at short distances and long times, a quasi-steady correlation pattern is present. This pattern is related to the excitations seen in $\sigma_\epsilon(\omega, \Delta t)$ at the excitonic resonances.

Let us now discuss the nature of the optical excitations and their dependence on $V$. For $V < W/2$, the correlation $C_{\rm dh}(r, t)$ shows that the doublon and holons are practically uncorrelated at all distances. This implies that there is no particular tendency of the system to develop bound optical excitations composed of tightly coupled doublon-holon pairs. Consequently, no resonances are expected to appear in $\sigma_\epsilon(\omega, \Delta t)$.

The situation is radically different for $V > W/2$, where distinct and diverse correlation patterns emerge. In particular, for $V = 3$, a clear tendency towards the pattern doublon-holon-doublon-holon is observed. This particular excitation corresponds to a (bi)exciton~\cite{Essler01, Jeck03} and is the main driver of the optical excitations of the electron system after interacting with the pump pulse seen in $\sigma_\epsilon(\omega, \Delta t)$. If we now consider the case $V = 4.5$, the correlation pattern exhibited by $C_{\rm dh}(r, t)$ is a `repeated' doublon-holon pattern and is much more structured and spans over larger distances than the $V = 3$ case. These optical excitations are excitonic strings,~\cite{Jeck03} which in this case are composed of approximately four tightly bound doublon-holon pairs.

In addition to the short distance patterns registered in $C_{\rm dh}(r, t)$, at large distances we notice that the correlation is not zero, although it certainly decays while oscillating between positive and negative values. This residual large-distance correlations are not present in the equilibrium optical excitations.

\begin{figure}
\includegraphics*[width=.35\textwidth]{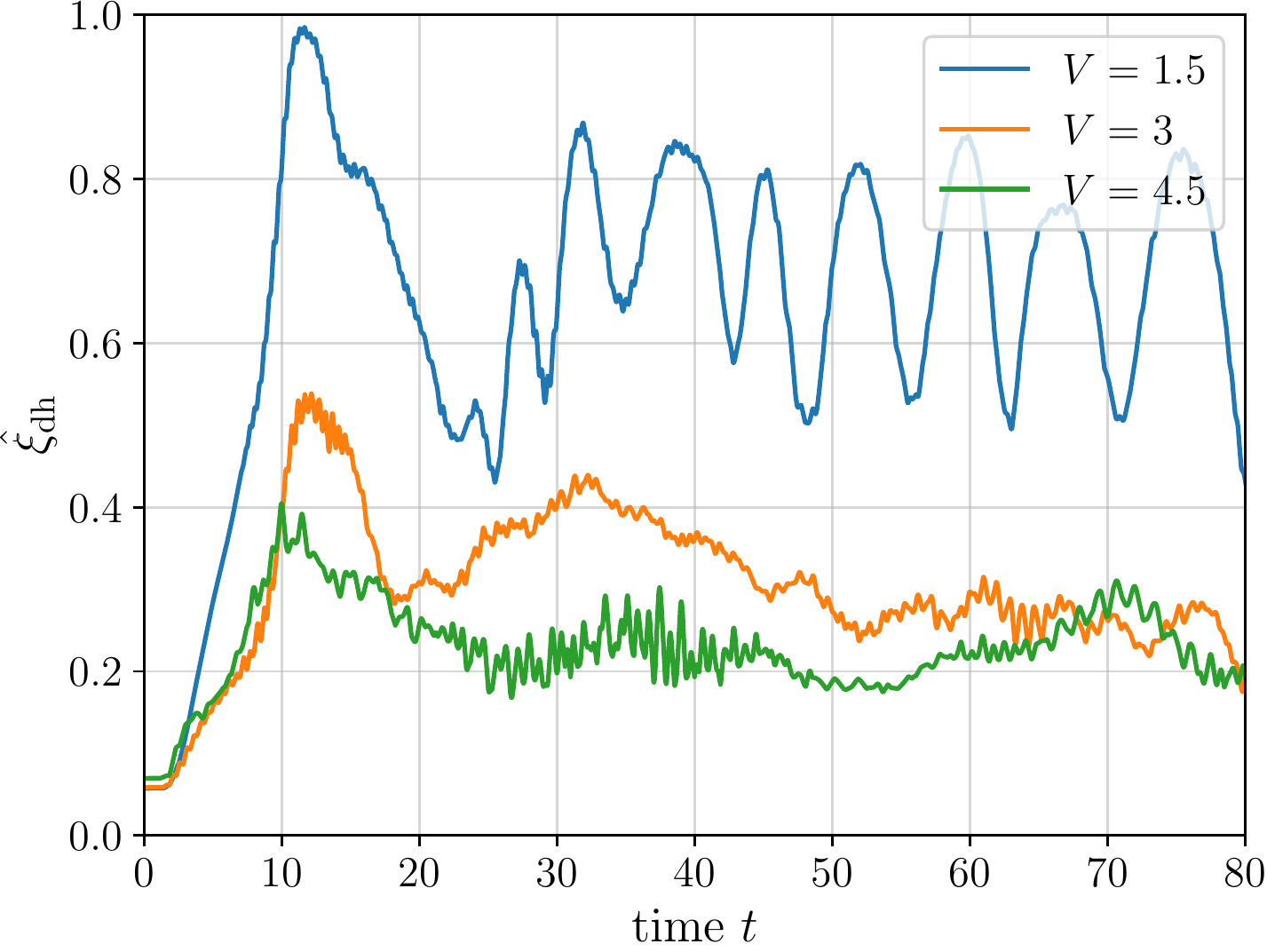}
\caption{Dimensionless average doublon-holon excitation size $\hat \xi_{\rm dh}$ as a function of time $t$ for the values of $V$ shown in Figs.~\ref{fig:sigmat} and \ref{fig:dhcorrt}. Values of $\hat \xi_{\rm dh} \ll 1$ indicate a tendency to form a bound excitation ($V = 3,\, 4.5$). Values of $\hat \xi_{\rm dh} \lesssim 1$ mark a tendency towards an unbound excitation ($V = 1.5$).}
\label{fig:xidh}
\end{figure}

\subsection{Optical excitations size}
We can further characterize the photoexcitations detected in $\sigma_\epsilon(\omega, \Delta t)$ with the average doublon-holon distance $\xi_{\rm dh}$, which is defined as the following weighted average of doublon-holon correlations:~\cite{Essler01, Jeck03}
\begin{equation}
\xi_{\rm dh}(t) = \frac{\sum_r |r| C_{\rm dh}(r, t)}{\sum_r C_{\rm dh}(r, t)} =: \frac{L}{2}\hat \xi_{\rm dh},
\end{equation}
where the sum goes from $r = 1$ to $L - 1$. If the doublon-holon excitation tends to be bound, $\xi_{\rm dh} \ll L/2$; on the other hand, if the tendency is opposite the doublon-holon pair is not bound and $\xi_{\rm dh} \sim L/2$. We introduce the dimensionless average doublon-holon distance
$
\hat \xi_{\rm dh} = \frac{2}{L} \xi_{\rm dh};
$ 
hence, for bound doublon-holon pairs $\hat \xi_{\rm dh} \ll 1$ and for unbound doublon-holon pairs $\hat \xi_{\rm dh} \lesssim 1$. Notice that the connected correlation function $C_{\rm dh}(r, t)$ decays to zero in equilibrium $t = 0$ as $r \to \infty$, therefore, providing a good measure for the calculation of the size of the excitations.

Figure~\ref{fig:xidh} displays the dimensionless average doublon-holon distance $\hat \xi_{\rm dh}(t)$ for values of $V$ discussed in Figs.~\ref{fig:s2}-\ref{fig:dhcorrt}. We notice that for $V < W/2$, the long-time average distance between doublon and holon fluctuates around $\hat \xi_{\rm dh} \approx 0.7 \lesssim 1$. We interpret this result as an indication that the correlation pattern observed in $C_{\rm dh}(r, t)$ describes holons and doublons behaving as independent excitations, since $\xi_{\rm dh} \lesssim L/2$. In contrast, for $V > W/2$, the doublon and holon are much more confined since $\hat \xi_{\rm dh} \approx 0.25 \ll 1$, and $\xi_{\rm dh} \ll L/2$. In this case, this result is indicative that the correlation pattern developed in $C_{\rm dh}(r, t)$ correspond to bound optical excitations.

\section{Conclusions}
We have examined the nonequilibrium optical response of a 1D Mott insulator to a pump pulse, in the particular context of the extended Hubbard model. We have derived an expression for the differential optical conductivity, which is related to the time-resolved optical conductivity: a key quantity in pump and probe spectroscopy experiments. We have computed it using a numerical prescription proposed in Ref.~\onlinecite{Shao:2016aa} and time-dependent DMRG. We detect a reduction of the local magnetic moment and concurrent photoexcitation of doublons and holons. The differential optical conductivity exhibits two main features: $(i)$ photogeneration of mid-gap spectral weight associated to parity-forbidden optical states, and $(ii)$ melting of the excitonic peak and emergence of a Fano optical resonance due to quantum interference of excitons and absorption band. The resulting nonequilibrium optical excitations are renormalized excitonic strings, (bi)excitons, or unbound doublon-holon pairs, upon decreasing of intersite Coulomb repulsion. 
Our results have direct relevance to pump and probe spectroscopy experiments in the THz domain performed on organic salts such as ET-F$_2$TCNQ, where quantum interference between excitons and absorption-band states have given rise the nonequilibrium optical excitations and photometallization.

\begin{acknowledgments}
AEF is supported by the U.S. Department of Energy, Office of Basic Energy Sciences under grant No.\ DE-SC0014407. JR thanks M. Zapata for fruitful discussions.
\end{acknowledgments}

\appendix

\section{Fourier transform and window functions\label{sec:fft}}
Given a signal $f(t)$, its Fourier transform is defined as
\begin{equation}
f(\omega) = \int_{- \infty}^{+ \infty} dt\, e^{i\omega t} \, W(t - t_0) f(t),
\label{eq:FT}
\end{equation}
where we have introduced the filter or window function $W(t - t_0)$. Useful window functions are
\begin{equation}
W(t - t_0) = 
\begin{cases}
\exp[- (t - t_0)^2 / 2 w^2] & \quad \textrm{Gauss}\\
\exp(- \eta |t - t_0|) & \quad \textrm{Poisson}\\
\cos^2[\pi (t - t_0) / c] & \quad \textrm{Hann}
\end{cases},
\end{equation}
where $\eta$, $w$, and $c$ are selected depending on the strength and robustness of the signal $f(t)$ at long times. For all practical purposes, all the simulations have a finite simulation time $t_{\rm DMRG}$ so that the Fourier transform~(\ref{eq:FT}) is performed in the interval $t \in [0, t_{\rm DMRG}]$ instead.

The time $t_0$ is the center of the window function. In time-resolved spectroscopy this will account for the time at which maximum amplitude of the electric field $E_{\rm pmp}(t)$ is reached. Before and after the peak of $E_{\rm pmp}(t)$ the current is not very much disturbed by it. The largest change to the current occurs around $t_0 = t_{\textrm{pmp}}$ so we centered the window function around it. The physical intuition behind using window functions is that $W(t - t_0)$ will partially account for decay and dephasing processes like electron-phonon and inelastic electron-electron interactions. 

In our calculations we have found similar results for different window functions. The chosen value for the window width, $w = 18\tau_{\rm pmp}$, offers a fair tradeoff between a weighted time average with high frequency resolution, while controlling the appearance of spurious poles, and a low impact of DMRG's truncation error at long times due to a finite-bond dimension MPS.

\begin{widetext}
\section{Relation to linear response theory\label{sec:linear}}
We derive a correspondence between the differential optical conductivity~(\ref{eq:sigma}) and a nonequilibrium form of the optical conductivity obtained from linear response theory. In the following, we neglect the diamagnetic term of the current operator to simplify the algebra.

We start from the equilibrium, linear-response expression of the current operator to an external perturbation (we use a hat to distinguish operators from expectation values)
\begin{equation}
J(t) = J_0 + i \int_{t_0}^t ds\, \braket{\Psi(0) | [\hat V(s), \hat J(t)] | \Psi(0)} + O(V^2).
\end{equation}
This expression assumes that we start with a many-body Hamiltonian $\hat H$ and the perturbation $\hat V(t)$ is switched on for $t > t_0$. $\hat J(t)$ is the current operator, and $J(t)$ and $J_0$ are its expectation values taken over $\ket{\Psi(t)}$ and $\ket{\Psi(0)}$, for the perturbed and unperturbed cases, respectively.

Now, we generalize that expression to a non-equilibrium pump-probe scenario, using the following protocol. First, the perturbation is due to the probe, such that $\hat V(t) = -\hat J(t) A(t)$, where $A(t) = A_{\rm prb}(t)$ is the probe pulse and $t_0 = t_{\rm prb}$ is the time at which we probe the system with an electric field $E(t) = -\partial_t A(t)$. (We drop the `prb' label for simplicity and reintroduce it in the final result.) Second, evolve the ground state, $\ket{\Psi(t = -\infty)}$, up to time $t_0$: $\ket{\Psi(t_0)} = \mathcal T \exp({-i \int_{-\infty}^{t_0} \hat H(\tau) d\tau})\ket{\Psi(t = -\infty)}$, where $\hat H(t)$ includes the pump pulse $A_{\rm pmp}(t)$, but not the probe field. Finally, at time $t_0$, the probe field is switched on. The resulting non-equilibrium, linear response of the system can be written as
\begin{equation}
J(t, t_0) = J_0(t) - i \int_{-\infty}^t ds\, \braket{\Psi(t_0) | [\hat J(s), \hat J(t)] | \Psi(t_0)} A(s) + O(A^2),
\end{equation}
where, to further simplify the expressions, we have introduced the shorthand notation $J(t, t_0) \leftarrow J(t, E, t_0)$, $J_0(t) \leftarrow J(t, E, 0)$. The expectation value $J(t, E, t_0)$ was defined at the beginning of Sec.~\ref{sec:diff_sigma}. Fourier transforming $J(t, t_0)$, $J_0(t)$, $A(s)$, and performing the change of variables $s \leftarrow t - s$ we obtain
\begin{equation}
J(\omega, t_0) = J_0(\omega) - i \int_0^{\infty} ds\, \braket{\Psi(t_0) | [\hat J(t - s), \hat J(t)] | \Psi(t_0)} e^{i \omega s} A(\omega) + O(A^2).
\end{equation}

Using the fact that $\hat J(t)$ is written in the interaction representation, $\hat J(t) = e^{i \hat H t} \hat J e^{-i \hat H t}$, and that in frequency domain the electric field and the vector potential satisfy $E(\omega) = i \omega A(\omega)$, we can further simplify to
\begin{align}
J(\omega, t_0) &= J_0(\omega) + \frac{1}{\omega} \int_0^{\infty} ds\, e^{i \omega s} \braket{\Psi(t_0) | [\hat J(s), \hat J(0)] | \Psi(t_0)} E(\omega), \\
&=: J_0(\omega) + \sigma(\omega, t_0) E(\omega),
\end{align}
where we have implicitly introduced the non-equilibrium optical conductivity $\sigma(\omega, t_0)$ and dropped any nonlinear contributions to $J(\omega, t_0)$. The definition $\sigma(\omega, t_0)$ is a non-equilibrium generalization of the familiar expression obtained from the Kubo formula.~\cite{Lenarcic2014, Shao:2016aa}

Starting from the above linear-response expression out of equilibrium, replacing the shorthand notation introduced before: $J(\omega, E, t_0) \leftarrow J(\omega, t_0)$, $J(\omega, E, 0) \leftarrow J_0(\omega)$, and reintroducing the label `prb,' the resulting expression for the expectation value of the current operator reads
\begin{equation}
J(\omega, E, t_{\rm prb}) = J(\omega, E, 0) + \sigma(\omega, t_{\rm prb}) E_{\rm prb}(\omega).
\end{equation}
Integrating with respect to $\omega \in [0, \infty)$, for fixed $E$ and $t_{\rm prb}$, and introducing the integrated current functional $\mathcal J(E)$~(\ref{eq:func}), we find that
\begin{equation}
\mathcal J(E) = \mathcal J_0(E) + \int_0^\infty d\omega\, \sigma(\omega, t_{\rm prb}) E_{\rm prb}(\omega).
\end{equation}
The functional differential of $\mathcal J$ is defined as $\delta \mathcal J := \mathcal J(E) - \mathcal J_0(E)$, such that the resulting expression is the same as Eq.~(\ref{eq:diff}), derived in Sec.~\ref{sec:diff_sigma}. From this result, we can reach the definition of the differential optical conductivity~(\ref{eq:sigma}), using the construction outlined in Sec.~\ref{sec:diff_sigma}.

We, therefore, have shown that starting from a non-equilibrium linear response theory it is possible to obtain the definition of the differential optical conductivity~(\ref{eq:sigma}), using the concept of functional derivative from calculus of variations.~\cite{Ewing2016}
\end{widetext}

\bibliographystyle{apsrev4-1}
\bibliography{sigma}

\begin{thebibliography}{52}%
\makeatletter
\providecommand \@ifxundefined [1]{%
 \@ifx{#1\undefined}
}%
\providecommand \@ifnum [1]{%
 \ifnum #1\expandafter \@firstoftwo
 \else \expandafter \@secondoftwo
 \fi
}%
\providecommand \@ifx [1]{%
 \ifx #1\expandafter \@firstoftwo
 \else \expandafter \@secondoftwo
 \fi
}%
\providecommand \natexlab [1]{#1}%
\providecommand \enquote  [1]{``#1''}%
\providecommand \bibnamefont  [1]{#1}%
\providecommand \bibfnamefont [1]{#1}%
\providecommand \citenamefont [1]{#1}%
\providecommand \href@noop [0]{\@secondoftwo}%
\providecommand \href [0]{\begingroup \@sanitize@url \@href}%
\providecommand \@href[1]{\@@startlink{#1}\@@href}%
\providecommand \@@href[1]{\endgroup#1\@@endlink}%
\providecommand \@sanitize@url [0]{\catcode `\\12\catcode `\$12\catcode
  `\&12\catcode `\#12\catcode `\^12\catcode `\_12\catcode `\%12\relax}%
\providecommand \@@startlink[1]{}%
\providecommand \@@endlink[0]{}%
\providecommand \url  [0]{\begingroup\@sanitize@url \@url }%
\providecommand \@url [1]{\endgroup\@href {#1}{\urlprefix }}%
\providecommand \urlprefix  [0]{URL }%
\providecommand \Eprint [0]{\href }%
\providecommand \doibase [0]{http://dx.doi.org/}%
\providecommand \selectlanguage [0]{\@gobble}%
\providecommand \bibinfo  [0]{\@secondoftwo}%
\providecommand \bibfield  [0]{\@secondoftwo}%
\providecommand \translation [1]{[#1]}%
\providecommand \BibitemOpen [0]{}%
\providecommand \bibitemStop [0]{}%
\providecommand \bibitemNoStop [0]{.\EOS\space}%
\providecommand \EOS [0]{\spacefactor3000\relax}%
\providecommand \BibitemShut  [1]{\csname bibitem#1\endcsname}%
\let\auto@bib@innerbib\@empty
\bibitem [{\citenamefont {Orenstein}(2012)}]{Orenstein_2012}%
  \BibitemOpen
  \bibfield  {author} {\bibinfo {author} {\bibfnamefont {J.}~\bibnamefont
  {Orenstein}},\ }\href {\doibase 10.1063/pt.3.1717} {\bibfield  {journal}
  {\bibinfo  {journal} {Physics Today}\ }\textbf {\bibinfo {volume} {65}},\
  \bibinfo {pages} {44} (\bibinfo {year} {2012})}\BibitemShut {NoStop}%
\bibitem [{\citenamefont {Kampfrath}\ \emph {et~al.}(2013)\citenamefont
  {Kampfrath}, \citenamefont {Tanaka},\ and\ \citenamefont
  {Nelson}}]{Kampfrath_2013}%
  \BibitemOpen
  \bibfield  {author} {\bibinfo {author} {\bibfnamefont {T.}~\bibnamefont
  {Kampfrath}}, \bibinfo {author} {\bibfnamefont {K.}~\bibnamefont {Tanaka}}, \
  and\ \bibinfo {author} {\bibfnamefont {K.~A.}\ \bibnamefont {Nelson}},\
  }\href {\doibase 10.1038/nphoton.2013.184} {\bibfield  {journal} {\bibinfo
  {journal} {Nature Photonics}\ }\textbf {\bibinfo {volume} {7}},\ \bibinfo
  {pages} {680} (\bibinfo {year} {2013})}\BibitemShut {NoStop}%
\bibitem [{\citenamefont {Zhang}\ and\ \citenamefont
  {Averitt}(2014)}]{Zhang_2014}%
  \BibitemOpen
  \bibfield  {author} {\bibinfo {author} {\bibfnamefont {J.}~\bibnamefont
  {Zhang}}\ and\ \bibinfo {author} {\bibfnamefont {R.}~\bibnamefont
  {Averitt}},\ }\href {\doibase 10.1146/annurev-matsci-070813-113258}
  {\bibfield  {journal} {\bibinfo  {journal} {Annual Review of Materials
  Research}\ }\textbf {\bibinfo {volume} {44}},\ \bibinfo {pages} {19}
  (\bibinfo {year} {2014})}\BibitemShut {NoStop}%
\bibitem [{\citenamefont {Gandolfi}\ \emph {et~al.}(2017)\citenamefont
  {Gandolfi}, \citenamefont {Celardo}, \citenamefont {Borgonovi}, \citenamefont
  {Ferrini}, \citenamefont {Avella}, \citenamefont {Banfi},\ and\ \citenamefont
  {Giannetti}}]{Gandolfi_2017}%
  \BibitemOpen
  \bibfield  {author} {\bibinfo {author} {\bibfnamefont {M.}~\bibnamefont
  {Gandolfi}}, \bibinfo {author} {\bibfnamefont {G.~L.}\ \bibnamefont
  {Celardo}}, \bibinfo {author} {\bibfnamefont {F.}~\bibnamefont {Borgonovi}},
  \bibinfo {author} {\bibfnamefont {G.}~\bibnamefont {Ferrini}}, \bibinfo
  {author} {\bibfnamefont {A.}~\bibnamefont {Avella}}, \bibinfo {author}
  {\bibfnamefont {F.}~\bibnamefont {Banfi}}, \ and\ \bibinfo {author}
  {\bibfnamefont {C.}~\bibnamefont {Giannetti}},\ }\href {\doibase
  10.1088/1402-4896/aa54cc} {\bibfield  {journal} {\bibinfo  {journal} {Physica
  Scripta}\ }\textbf {\bibinfo {volume} {92}},\ \bibinfo {pages} {034004}
  (\bibinfo {year} {2017})}\BibitemShut {NoStop}%
\bibitem [{\citenamefont {Basov}\ \emph {et~al.}(2017)\citenamefont {Basov},
  \citenamefont {Averitt},\ and\ \citenamefont {Hsieh}}]{Basov_2017}%
  \BibitemOpen
  \bibfield  {author} {\bibinfo {author} {\bibfnamefont {D.~N.}\ \bibnamefont
  {Basov}}, \bibinfo {author} {\bibfnamefont {R.~D.}\ \bibnamefont {Averitt}},
  \ and\ \bibinfo {author} {\bibfnamefont {D.}~\bibnamefont {Hsieh}},\ }\href
  {\doibase 10.1038/nmat5017} {\bibfield  {journal} {\bibinfo  {journal}
  {Nature Materials}\ }\textbf {\bibinfo {volume} {16}},\ \bibinfo {pages}
  {1077} (\bibinfo {year} {2017})}\BibitemShut {NoStop}%
\bibitem [{\citenamefont {Wang}\ \emph {et~al.}(2018)\citenamefont {Wang},
  \citenamefont {Claassen}, \citenamefont {Pemmaraju}, \citenamefont {Jia},
  \citenamefont {Moritz},\ and\ \citenamefont {Devereaux}}]{Wang_2018}%
  \BibitemOpen
  \bibfield  {author} {\bibinfo {author} {\bibfnamefont {Y.}~\bibnamefont
  {Wang}}, \bibinfo {author} {\bibfnamefont {M.}~\bibnamefont {Claassen}},
  \bibinfo {author} {\bibfnamefont {C.~D.}\ \bibnamefont {Pemmaraju}}, \bibinfo
  {author} {\bibfnamefont {C.}~\bibnamefont {Jia}}, \bibinfo {author}
  {\bibfnamefont {B.}~\bibnamefont {Moritz}}, \ and\ \bibinfo {author}
  {\bibfnamefont {T.~P.}\ \bibnamefont {Devereaux}},\ }\href {\doibase
  10.1038/s41578-018-0046-3} {\bibfield  {journal} {\bibinfo  {journal} {Nature
  Reviews Materials}\ }\textbf {\bibinfo {volume} {3}},\ \bibinfo {pages} {312}
  (\bibinfo {year} {2018})}\BibitemShut {NoStop}%
\bibitem [{\citenamefont {Wall}\ \emph {et~al.}(2010)\citenamefont {Wall},
  \citenamefont {Brida}, \citenamefont {Clark}, \citenamefont {Ehrke},
  \citenamefont {Jaksch}, \citenamefont {Ardavan}, \citenamefont {Bonora},
  \citenamefont {Uemura}, \citenamefont {Takahashi}, \citenamefont {Hasegawa},
  \citenamefont {Okamoto}, \citenamefont {Cerullo},\ and\ \citenamefont
  {Cavalleri}}]{Wall_2010}%
  \BibitemOpen
  \bibfield  {author} {\bibinfo {author} {\bibfnamefont {S.}~\bibnamefont
  {Wall}}, \bibinfo {author} {\bibfnamefont {D.}~\bibnamefont {Brida}},
  \bibinfo {author} {\bibfnamefont {S.~R.}\ \bibnamefont {Clark}}, \bibinfo
  {author} {\bibfnamefont {H.~P.}\ \bibnamefont {Ehrke}}, \bibinfo {author}
  {\bibfnamefont {D.}~\bibnamefont {Jaksch}}, \bibinfo {author} {\bibfnamefont
  {A.}~\bibnamefont {Ardavan}}, \bibinfo {author} {\bibfnamefont
  {S.}~\bibnamefont {Bonora}}, \bibinfo {author} {\bibfnamefont
  {H.}~\bibnamefont {Uemura}}, \bibinfo {author} {\bibfnamefont
  {Y.}~\bibnamefont {Takahashi}}, \bibinfo {author} {\bibfnamefont
  {T.}~\bibnamefont {Hasegawa}}, \bibinfo {author} {\bibfnamefont
  {H.}~\bibnamefont {Okamoto}}, \bibinfo {author} {\bibfnamefont
  {G.}~\bibnamefont {Cerullo}}, \ and\ \bibinfo {author} {\bibfnamefont
  {A.}~\bibnamefont {Cavalleri}},\ }\href {\doibase 10.1038/nphys1831}
  {\bibfield  {journal} {\bibinfo  {journal} {Nature Physics}\ }\textbf
  {\bibinfo {volume} {7}},\ \bibinfo {pages} {114} (\bibinfo {year}
  {2010})}\BibitemShut {NoStop}%
\bibitem [{\citenamefont {Mitrano}\ \emph {et~al.}(2014)\citenamefont
  {Mitrano}, \citenamefont {Cotugno}, \citenamefont {Clark}, \citenamefont
  {Singla}, \citenamefont {Kaiser}, \citenamefont {St\"ahler}, \citenamefont
  {Beyer}, \citenamefont {Dressel}, \citenamefont {Baldassarre}, \citenamefont
  {Nicoletti}, \citenamefont {Perucchi}, \citenamefont {Hasegawa},
  \citenamefont {Okamoto}, \citenamefont {Jaksch},\ and\ \citenamefont
  {Cavalleri}}]{Mitrano14}%
  \BibitemOpen
  \bibfield  {author} {\bibinfo {author} {\bibfnamefont {M.}~\bibnamefont
  {Mitrano}}, \bibinfo {author} {\bibfnamefont {G.}~\bibnamefont {Cotugno}},
  \bibinfo {author} {\bibfnamefont {S.~R.}\ \bibnamefont {Clark}}, \bibinfo
  {author} {\bibfnamefont {R.}~\bibnamefont {Singla}}, \bibinfo {author}
  {\bibfnamefont {S.}~\bibnamefont {Kaiser}}, \bibinfo {author} {\bibfnamefont
  {J.}~\bibnamefont {St\"ahler}}, \bibinfo {author} {\bibfnamefont
  {R.}~\bibnamefont {Beyer}}, \bibinfo {author} {\bibfnamefont
  {M.}~\bibnamefont {Dressel}}, \bibinfo {author} {\bibfnamefont
  {L.}~\bibnamefont {Baldassarre}}, \bibinfo {author} {\bibfnamefont
  {D.}~\bibnamefont {Nicoletti}}, \bibinfo {author} {\bibfnamefont
  {A.}~\bibnamefont {Perucchi}}, \bibinfo {author} {\bibfnamefont
  {T.}~\bibnamefont {Hasegawa}}, \bibinfo {author} {\bibfnamefont
  {H.}~\bibnamefont {Okamoto}}, \bibinfo {author} {\bibfnamefont
  {D.}~\bibnamefont {Jaksch}}, \ and\ \bibinfo {author} {\bibfnamefont
  {A.}~\bibnamefont {Cavalleri}},\ }\href {\doibase
  10.1103/PhysRevLett.112.117801} {\bibfield  {journal} {\bibinfo  {journal}
  {Physical Review Letters}\ }\textbf {\bibinfo {volume} {112}},\ \bibinfo
  {pages} {117801} (\bibinfo {year} {2014})}\BibitemShut {NoStop}%
\bibitem [{\citenamefont {Miyamoto}\ \emph {et~al.}(2019)\citenamefont
  {Miyamoto}, \citenamefont {Kakizaki}, \citenamefont {Terashige},
  \citenamefont {Hata}, \citenamefont {Yamakawa}, \citenamefont {Morimoto},
  \citenamefont {Takamura}, \citenamefont {Yada}, \citenamefont {Takahashi},
  \citenamefont {Hasegawa}, \citenamefont {Matsuzaki}, \citenamefont
  {Tohyama},\ and\ \citenamefont {Okamoto}}]{Miyamoto_2019}%
  \BibitemOpen
  \bibfield  {author} {\bibinfo {author} {\bibfnamefont {T.}~\bibnamefont
  {Miyamoto}}, \bibinfo {author} {\bibfnamefont {T.}~\bibnamefont {Kakizaki}},
  \bibinfo {author} {\bibfnamefont {T.}~\bibnamefont {Terashige}}, \bibinfo
  {author} {\bibfnamefont {D.}~\bibnamefont {Hata}}, \bibinfo {author}
  {\bibfnamefont {H.}~\bibnamefont {Yamakawa}}, \bibinfo {author}
  {\bibfnamefont {T.}~\bibnamefont {Morimoto}}, \bibinfo {author}
  {\bibfnamefont {N.}~\bibnamefont {Takamura}}, \bibinfo {author}
  {\bibfnamefont {H.}~\bibnamefont {Yada}}, \bibinfo {author} {\bibfnamefont
  {Y.}~\bibnamefont {Takahashi}}, \bibinfo {author} {\bibfnamefont
  {T.}~\bibnamefont {Hasegawa}}, \bibinfo {author} {\bibfnamefont
  {H.}~\bibnamefont {Matsuzaki}}, \bibinfo {author} {\bibfnamefont
  {T.}~\bibnamefont {Tohyama}}, \ and\ \bibinfo {author} {\bibfnamefont
  {H.}~\bibnamefont {Okamoto}},\ }\href {\doibase 10.1038/s42005-019-0223-8}
  {\bibfield  {journal} {\bibinfo  {journal} {Communications Physics}\ }\textbf
  {\bibinfo {volume} {2}},\ \bibinfo {pages} {131} (\bibinfo {year}
  {2019})}\BibitemShut {NoStop}%
\bibitem [{\citenamefont {Ono}\ \emph {et~al.}(2005)\citenamefont {Ono},
  \citenamefont {Kishida},\ and\ \citenamefont {Okamoto}}]{Ono:2005aa}%
  \BibitemOpen
  \bibfield  {author} {\bibinfo {author} {\bibfnamefont {M.}~\bibnamefont
  {Ono}}, \bibinfo {author} {\bibfnamefont {H.}~\bibnamefont {Kishida}}, \ and\
  \bibinfo {author} {\bibfnamefont {H.}~\bibnamefont {Okamoto}},\ }\href
  {\doibase 10.1103/PhysRevLett.95.087401} {\bibfield  {journal} {\bibinfo
  {journal} {Physical Review Letters}\ }\textbf {\bibinfo {volume} {95}},\
  \bibinfo {pages} {087401} (\bibinfo {year} {2005})}\BibitemShut {NoStop}%
\bibitem [{\citenamefont {Okamoto}\ \emph {et~al.}(2007)\citenamefont
  {Okamoto}, \citenamefont {Matsuzaki}, \citenamefont {Wakabayashi},
  \citenamefont {Takahashi},\ and\ \citenamefont {Hasegawa}}]{Okamoto07}%
  \BibitemOpen
  \bibfield  {author} {\bibinfo {author} {\bibfnamefont {H.}~\bibnamefont
  {Okamoto}}, \bibinfo {author} {\bibfnamefont {H.}~\bibnamefont {Matsuzaki}},
  \bibinfo {author} {\bibfnamefont {T.}~\bibnamefont {Wakabayashi}}, \bibinfo
  {author} {\bibfnamefont {Y.}~\bibnamefont {Takahashi}}, \ and\ \bibinfo
  {author} {\bibfnamefont {T.}~\bibnamefont {Hasegawa}},\ }\href {\doibase
  10.1103/PhysRevLett.98.037401} {\bibfield  {journal} {\bibinfo  {journal}
  {Physical Review Letters}\ }\textbf {\bibinfo {volume} {98}},\ \bibinfo
  {pages} {037401} (\bibinfo {year} {2007})}\BibitemShut {NoStop}%
\bibitem [{\citenamefont {Yamakawa}\ \emph {et~al.}(2017)\citenamefont
  {Yamakawa}, \citenamefont {Miyamoto}, \citenamefont {Morimoto}, \citenamefont
  {Terashige}, \citenamefont {Yada}, \citenamefont {Kida}, \citenamefont
  {Suda}, \citenamefont {Yamamoto}, \citenamefont {Kato}, \citenamefont
  {Miyagawa}, \citenamefont {Kanoda},\ and\ \citenamefont
  {Okamoto}}]{Yamakawa_2017}%
  \BibitemOpen
  \bibfield  {author} {\bibinfo {author} {\bibfnamefont {H.}~\bibnamefont
  {Yamakawa}}, \bibinfo {author} {\bibfnamefont {T.}~\bibnamefont {Miyamoto}},
  \bibinfo {author} {\bibfnamefont {T.}~\bibnamefont {Morimoto}}, \bibinfo
  {author} {\bibfnamefont {T.}~\bibnamefont {Terashige}}, \bibinfo {author}
  {\bibfnamefont {H.}~\bibnamefont {Yada}}, \bibinfo {author} {\bibfnamefont
  {N.}~\bibnamefont {Kida}}, \bibinfo {author} {\bibfnamefont {M.}~\bibnamefont
  {Suda}}, \bibinfo {author} {\bibfnamefont {H.~M.}\ \bibnamefont {Yamamoto}},
  \bibinfo {author} {\bibfnamefont {R.}~\bibnamefont {Kato}}, \bibinfo {author}
  {\bibfnamefont {K.}~\bibnamefont {Miyagawa}}, \bibinfo {author}
  {\bibfnamefont {K.}~\bibnamefont {Kanoda}}, \ and\ \bibinfo {author}
  {\bibfnamefont {H.}~\bibnamefont {Okamoto}},\ }\href {\doibase
  10.1038/nmat4967} {\bibfield  {journal} {\bibinfo  {journal} {Nature
  Materials}\ }\textbf {\bibinfo {volume} {16}},\ \bibinfo {pages} {1100}
  (\bibinfo {year} {2017})}\BibitemShut {NoStop}%
\bibitem [{\citenamefont {Frenzel}\ \emph {et~al.}(2013)\citenamefont
  {Frenzel}, \citenamefont {Lui}, \citenamefont {Fang}, \citenamefont {Nair},
  \citenamefont {Herring}, \citenamefont {Jarillo-Herrero}, \citenamefont
  {Kong},\ and\ \citenamefont {Gedik}}]{Frenzel_2013}%
  \BibitemOpen
  \bibfield  {author} {\bibinfo {author} {\bibfnamefont {A.~J.}\ \bibnamefont
  {Frenzel}}, \bibinfo {author} {\bibfnamefont {C.~H.}\ \bibnamefont {Lui}},
  \bibinfo {author} {\bibfnamefont {W.}~\bibnamefont {Fang}}, \bibinfo {author}
  {\bibfnamefont {N.~L.}\ \bibnamefont {Nair}}, \bibinfo {author}
  {\bibfnamefont {P.~K.}\ \bibnamefont {Herring}}, \bibinfo {author}
  {\bibfnamefont {P.}~\bibnamefont {Jarillo-Herrero}}, \bibinfo {author}
  {\bibfnamefont {J.}~\bibnamefont {Kong}}, \ and\ \bibinfo {author}
  {\bibfnamefont {N.}~\bibnamefont {Gedik}},\ }\href {\doibase
  10.1063/1.4795858} {\bibfield  {journal} {\bibinfo  {journal} {Applied
  Physics Letters}\ }\textbf {\bibinfo {volume} {102}},\ \bibinfo {pages}
  {113111} (\bibinfo {year} {2013})}\BibitemShut {NoStop}%
\bibitem [{\citenamefont {Lui}\ \emph {et~al.}(2014)\citenamefont {Lui},
  \citenamefont {Frenzel}, \citenamefont {Pilon}, \citenamefont {Lee},
  \citenamefont {Ling}, \citenamefont {Akselrod}, \citenamefont {Kong},\ and\
  \citenamefont {Gedik}}]{Lui:2014aa}%
  \BibitemOpen
  \bibfield  {author} {\bibinfo {author} {\bibfnamefont {C.}~\bibnamefont
  {Lui}}, \bibinfo {author} {\bibfnamefont {A.}~\bibnamefont {Frenzel}},
  \bibinfo {author} {\bibfnamefont {D.}~\bibnamefont {Pilon}}, \bibinfo
  {author} {\bibfnamefont {Y.-H.}\ \bibnamefont {Lee}}, \bibinfo {author}
  {\bibfnamefont {X.}~\bibnamefont {Ling}}, \bibinfo {author} {\bibfnamefont
  {G.}~\bibnamefont {Akselrod}}, \bibinfo {author} {\bibfnamefont
  {J.}~\bibnamefont {Kong}}, \ and\ \bibinfo {author} {\bibfnamefont
  {N.}~\bibnamefont {Gedik}},\ }\href {\doibase 10.1103/PhysRevLett.113.166801}
  {\bibfield  {journal} {\bibinfo  {journal} {Physical Review Letters}\
  }\textbf {\bibinfo {volume} {113}},\ \bibinfo {pages} {166801} (\bibinfo
  {year} {2014})}\BibitemShut {NoStop}%
\bibitem [{\citenamefont {Matsubara}\ \emph {et~al.}(2014)\citenamefont
  {Matsubara}, \citenamefont {Ogihara}, \citenamefont {Itatani}, \citenamefont
  {Maeshima}, \citenamefont {Yonemitsu}, \citenamefont {Ishikawa},
  \citenamefont {Okimoto}, \citenamefont {ya~Koshihara}, \citenamefont
  {Hiramatsu}, \citenamefont {Nakano}, \citenamefont {Yamochi}, \citenamefont
  {Saito},\ and\ \citenamefont {Onda}}]{Matsubara:2014aa}%
  \BibitemOpen
  \bibfield  {author} {\bibinfo {author} {\bibfnamefont {Y.}~\bibnamefont
  {Matsubara}}, \bibinfo {author} {\bibfnamefont {S.}~\bibnamefont {Ogihara}},
  \bibinfo {author} {\bibfnamefont {J.}~\bibnamefont {Itatani}}, \bibinfo
  {author} {\bibfnamefont {N.}~\bibnamefont {Maeshima}}, \bibinfo {author}
  {\bibfnamefont {K.}~\bibnamefont {Yonemitsu}}, \bibinfo {author}
  {\bibfnamefont {T.}~\bibnamefont {Ishikawa}}, \bibinfo {author}
  {\bibfnamefont {Y.}~\bibnamefont {Okimoto}}, \bibinfo {author} {\bibfnamefont
  {S.}~\bibnamefont {ya~Koshihara}}, \bibinfo {author} {\bibfnamefont
  {T.}~\bibnamefont {Hiramatsu}}, \bibinfo {author} {\bibfnamefont
  {Y.}~\bibnamefont {Nakano}}, \bibinfo {author} {\bibfnamefont
  {H.}~\bibnamefont {Yamochi}}, \bibinfo {author} {\bibfnamefont
  {G.}~\bibnamefont {Saito}}, \ and\ \bibinfo {author} {\bibfnamefont
  {K.}~\bibnamefont {Onda}},\ }\href {\doibase 10.1103/PhysRevB.89.161102}
  {\bibfield  {journal} {\bibinfo  {journal} {Physical Review B}\ }\textbf
  {\bibinfo {volume} {89}},\ \bibinfo {pages} {161102(R)} (\bibinfo {year}
  {2014})}\BibitemShut {NoStop}%
\bibitem [{\citenamefont {Matsuzaki}\ \emph {et~al.}(2006)\citenamefont
  {Matsuzaki}, \citenamefont {Yamashita},\ and\ \citenamefont
  {Okamoto}}]{Matsuzaki_2006}%
  \BibitemOpen
  \bibfield  {author} {\bibinfo {author} {\bibfnamefont {H.}~\bibnamefont
  {Matsuzaki}}, \bibinfo {author} {\bibfnamefont {M.}~\bibnamefont
  {Yamashita}}, \ and\ \bibinfo {author} {\bibfnamefont {H.}~\bibnamefont
  {Okamoto}},\ }\href {\doibase 10.1143/jpsj.75.123701} {\bibfield  {journal}
  {\bibinfo  {journal} {Journal of the Physical Society of Japan}\ }\textbf
  {\bibinfo {volume} {75}},\ \bibinfo {pages} {123701} (\bibinfo {year}
  {2006})}\BibitemShut {NoStop}%
\bibitem [{\citenamefont {Gebhard}\ \emph
  {et~al.}(1997{\natexlab{a}})\citenamefont {Gebhard}, \citenamefont {Bott},
  \citenamefont {Scheidler}, \citenamefont {Thomas},\ and\ \citenamefont
  {Koch}}]{Gebhard_1997a}%
  \BibitemOpen
  \bibfield  {author} {\bibinfo {author} {\bibfnamefont {F.}~\bibnamefont
  {Gebhard}}, \bibinfo {author} {\bibfnamefont {K.}~\bibnamefont {Bott}},
  \bibinfo {author} {\bibfnamefont {M.}~\bibnamefont {Scheidler}}, \bibinfo
  {author} {\bibfnamefont {P.}~\bibnamefont {Thomas}}, \ and\ \bibinfo {author}
  {\bibfnamefont {S.~W.}\ \bibnamefont {Koch}},\ }\href {\doibase
  10.1080/13642819708205700} {\bibfield  {journal} {\bibinfo  {journal}
  {Philosophical Magazine B}\ }\textbf {\bibinfo {volume} {75}},\ \bibinfo
  {pages} {1} (\bibinfo {year} {1997}{\natexlab{a}})}\BibitemShut {NoStop}%
\bibitem [{\citenamefont {Gebhard}\ \emph
  {et~al.}(1997{\natexlab{b}})\citenamefont {Gebhard}, \citenamefont {Born},
  \citenamefont {Scheidler}, \citenamefont {Thomas},\ and\ \citenamefont
  {Koch}}]{Gebhard_1997b}%
  \BibitemOpen
  \bibfield  {author} {\bibinfo {author} {\bibfnamefont {F.}~\bibnamefont
  {Gebhard}}, \bibinfo {author} {\bibfnamefont {K.}~\bibnamefont {Born}},
  \bibinfo {author} {\bibfnamefont {M.}~\bibnamefont {Scheidler}}, \bibinfo
  {author} {\bibfnamefont {P.}~\bibnamefont {Thomas}}, \ and\ \bibinfo {author}
  {\bibfnamefont {S.~W.}\ \bibnamefont {Koch}},\ }\href {\doibase
  10.1080/13642819708205701} {\bibfield  {journal} {\bibinfo  {journal}
  {Philosophical Magazine B}\ }\textbf {\bibinfo {volume} {75}},\ \bibinfo
  {pages} {13} (\bibinfo {year} {1997}{\natexlab{b}})}\BibitemShut {NoStop}%
\bibitem [{\citenamefont {Gebhard}\ \emph
  {et~al.}(1997{\natexlab{c}})\citenamefont {Gebhard}, \citenamefont {Born},
  \citenamefont {Scheidler}, \citenamefont {Thomas},\ and\ \citenamefont
  {Koch}}]{Gebhard_1997c}%
  \BibitemOpen
  \bibfield  {author} {\bibinfo {author} {\bibfnamefont {F.}~\bibnamefont
  {Gebhard}}, \bibinfo {author} {\bibfnamefont {K.}~\bibnamefont {Born}},
  \bibinfo {author} {\bibfnamefont {M.}~\bibnamefont {Scheidler}}, \bibinfo
  {author} {\bibfnamefont {P.}~\bibnamefont {Thomas}}, \ and\ \bibinfo {author}
  {\bibfnamefont {S.~W.}\ \bibnamefont {Koch}},\ }\href {\doibase
  10.1080/13642819708205702} {\bibfield  {journal} {\bibinfo  {journal}
  {Philosophical Magazine B}\ }\textbf {\bibinfo {volume} {75}},\ \bibinfo
  {pages} {47} (\bibinfo {year} {1997}{\natexlab{c}})}\BibitemShut {NoStop}%
\bibitem [{\citenamefont {Gallagher}\ and\ \citenamefont
  {Mazumdar}(1997)}]{Gallagher:1997aa}%
  \BibitemOpen
  \bibfield  {author} {\bibinfo {author} {\bibfnamefont {F.~B.}\ \bibnamefont
  {Gallagher}}\ and\ \bibinfo {author} {\bibfnamefont {S.}~\bibnamefont
  {Mazumdar}},\ }\href {\doibase 10.1103/PhysRevB.56.15025} {\bibfield
  {journal} {\bibinfo  {journal} {Physical Review B}\ }\textbf {\bibinfo
  {volume} {56}},\ \bibinfo {pages} {15025} (\bibinfo {year}
  {1997})}\BibitemShut {NoStop}%
\bibitem [{\citenamefont {Kancharla}\ and\ \citenamefont
  {Bolech}(2001)}]{Kancharla:2001aa}%
  \BibitemOpen
  \bibfield  {author} {\bibinfo {author} {\bibfnamefont {S.~S.}\ \bibnamefont
  {Kancharla}}\ and\ \bibinfo {author} {\bibfnamefont {C.~J.}\ \bibnamefont
  {Bolech}},\ }\href {\doibase 10.1103/PhysRevB.64.085119} {\bibfield
  {journal} {\bibinfo  {journal} {Physical Review B}\ }\textbf {\bibinfo
  {volume} {64}},\ \bibinfo {pages} {085119} (\bibinfo {year}
  {2001})}\BibitemShut {NoStop}%
\bibitem [{\citenamefont {Essler}\ \emph {et~al.}(2001)\citenamefont {Essler},
  \citenamefont {Gebhard},\ and\ \citenamefont {Jeckelmann}}]{Essler01}%
  \BibitemOpen
  \bibfield  {author} {\bibinfo {author} {\bibfnamefont {F.~H.~L.}\
  \bibnamefont {Essler}}, \bibinfo {author} {\bibfnamefont {F.}~\bibnamefont
  {Gebhard}}, \ and\ \bibinfo {author} {\bibfnamefont {E.}~\bibnamefont
  {Jeckelmann}},\ }\href {\doibase 10.1103/PhysRevB.64.125119} {\bibfield
  {journal} {\bibinfo  {journal} {Physical Review B}\ }\textbf {\bibinfo
  {volume} {64}},\ \bibinfo {pages} {125119} (\bibinfo {year}
  {2001})}\BibitemShut {NoStop}%
\bibitem [{\citenamefont {Jeckelmann}(2003)}]{Jeck03}%
  \BibitemOpen
  \bibfield  {author} {\bibinfo {author} {\bibfnamefont {E.}~\bibnamefont
  {Jeckelmann}},\ }\href {\doibase 10.1103/PhysRevB.67.075106} {\bibfield
  {journal} {\bibinfo  {journal} {Physical Review B}\ }\textbf {\bibinfo
  {volume} {67}},\ \bibinfo {pages} {075106} (\bibinfo {year}
  {2003})}\BibitemShut {NoStop}%
\bibitem [{\citenamefont {Al-Hassanieh}\ \emph {et~al.}(2008)\citenamefont
  {Al-Hassanieh}, \citenamefont {Reboredo}, \citenamefont {Feiguin},
  \citenamefont {Gonz{\'a}lez},\ and\ \citenamefont
  {Dagotto}}]{Al-Hassanieh:2008aa}%
  \BibitemOpen
  \bibfield  {author} {\bibinfo {author} {\bibfnamefont {K.~A.}\ \bibnamefont
  {Al-Hassanieh}}, \bibinfo {author} {\bibfnamefont {F.~A.}\ \bibnamefont
  {Reboredo}}, \bibinfo {author} {\bibfnamefont {A.~E.}\ \bibnamefont
  {Feiguin}}, \bibinfo {author} {\bibfnamefont {I.}~\bibnamefont
  {Gonz{\'a}lez}}, \ and\ \bibinfo {author} {\bibfnamefont {E.}~\bibnamefont
  {Dagotto}},\ }\href {\doibase 10.1103/PhysRevLett.100.166403} {\bibfield
  {journal} {\bibinfo  {journal} {Physical Review Letters}\ }\textbf {\bibinfo
  {volume} {100}},\ \bibinfo {pages} {166403} (\bibinfo {year}
  {2008})}\BibitemShut {NoStop}%
\bibitem [{\citenamefont {Lenar{\v c}i{\v c}}\ and\ \citenamefont {Prelov{\v
  s}ek}(2013)}]{Lenarcic:2013aa}%
  \BibitemOpen
  \bibfield  {author} {\bibinfo {author} {\bibfnamefont {Z.}~\bibnamefont
  {Lenar{\v c}i{\v c}}}\ and\ \bibinfo {author} {\bibfnamefont
  {P.}~\bibnamefont {Prelov{\v s}ek}},\ }\href {\doibase
  10.1103/PhysRevLett.111.016401} {\bibfield  {journal} {\bibinfo  {journal}
  {Physical Review Letters}\ }\textbf {\bibinfo {volume} {111}},\ \bibinfo
  {pages} {016401} (\bibinfo {year} {2013})}\BibitemShut {NoStop}%
\bibitem [{\citenamefont {Lenar{\v c}i{\v c}}\ \emph
  {et~al.}(2015)\citenamefont {Lenar{\v c}i{\v c}}, \citenamefont {Eckstein},\
  and\ \citenamefont {Prelov{\v s}ek}}]{Lenarcic:2015aa}%
  \BibitemOpen
  \bibfield  {author} {\bibinfo {author} {\bibfnamefont {Z.}~\bibnamefont
  {Lenar{\v c}i{\v c}}}, \bibinfo {author} {\bibfnamefont {M.}~\bibnamefont
  {Eckstein}}, \ and\ \bibinfo {author} {\bibfnamefont {P.}~\bibnamefont
  {Prelov{\v s}ek}},\ }\href {\doibase 10.1103/PhysRevB.92.201104} {\bibfield
  {journal} {\bibinfo  {journal} {Physical Review B}\ }\textbf {\bibinfo
  {volume} {92}},\ \bibinfo {pages} {201104(R)} (\bibinfo {year}
  {2015})}\BibitemShut {NoStop}%
\bibitem [{\citenamefont {Ohmura}\ \emph {et~al.}(2019)\citenamefont {Ohmura},
  \citenamefont {Takahashi}, \citenamefont {Iwano}, \citenamefont {Yamaguchi},
  \citenamefont {Shinjo}, \citenamefont {Tohyama}, \citenamefont {Sota},\ and\
  \citenamefont {Okamoto}}]{Ohmura:2019aa}%
  \BibitemOpen
  \bibfield  {author} {\bibinfo {author} {\bibfnamefont {S.}~\bibnamefont
  {Ohmura}}, \bibinfo {author} {\bibfnamefont {A.}~\bibnamefont {Takahashi}},
  \bibinfo {author} {\bibfnamefont {K.}~\bibnamefont {Iwano}}, \bibinfo
  {author} {\bibfnamefont {T.}~\bibnamefont {Yamaguchi}}, \bibinfo {author}
  {\bibfnamefont {K.}~\bibnamefont {Shinjo}}, \bibinfo {author} {\bibfnamefont
  {T.}~\bibnamefont {Tohyama}}, \bibinfo {author} {\bibfnamefont
  {S.}~\bibnamefont {Sota}}, \ and\ \bibinfo {author} {\bibfnamefont
  {H.}~\bibnamefont {Okamoto}},\ }\href {\doibase 10.1103/PhysRevB.100.235134}
  {\bibfield  {journal} {\bibinfo  {journal} {Physical Review B}\ }\textbf
  {\bibinfo {volume} {100}},\ \bibinfo {pages} {235134} (\bibinfo {year}
  {2019})}\BibitemShut {NoStop}%
\bibitem [{\citenamefont {Gole{\v z}}\ \emph {et~al.}(2015)\citenamefont
  {Gole{\v z}}, \citenamefont {Eckstein},\ and\ \citenamefont
  {Werner}}]{Golez:2015aa}%
  \BibitemOpen
  \bibfield  {author} {\bibinfo {author} {\bibfnamefont {D.}~\bibnamefont
  {Gole{\v z}}}, \bibinfo {author} {\bibfnamefont {M.}~\bibnamefont
  {Eckstein}}, \ and\ \bibinfo {author} {\bibfnamefont {P.}~\bibnamefont
  {Werner}},\ }\href {\doibase 10.1103/PhysRevB.92.195123} {\bibfield
  {journal} {\bibinfo  {journal} {Physical Review B}\ }\textbf {\bibinfo
  {volume} {92}},\ \bibinfo {pages} {195123} (\bibinfo {year}
  {2015})}\BibitemShut {NoStop}%
\bibitem [{\citenamefont {Rinc{\'o}n}\ \emph {et~al.}(2014)\citenamefont
  {Rinc{\'o}n}, \citenamefont {Al-Hassanieh}, \citenamefont {Feiguin},\ and\
  \citenamefont {Dagotto}}]{Rincon:2014aa}%
  \BibitemOpen
  \bibfield  {author} {\bibinfo {author} {\bibfnamefont {J.}~\bibnamefont
  {Rinc{\'o}n}}, \bibinfo {author} {\bibfnamefont {K.~A.}\ \bibnamefont
  {Al-Hassanieh}}, \bibinfo {author} {\bibfnamefont {A.~E.}\ \bibnamefont
  {Feiguin}}, \ and\ \bibinfo {author} {\bibfnamefont {E.}~\bibnamefont
  {Dagotto}},\ }\href {\doibase 10.1103/PhysRevB.90.155112} {\bibfield
  {journal} {\bibinfo  {journal} {Physical Review B}\ }\textbf {\bibinfo
  {volume} {90}},\ \bibinfo {pages} {155112} (\bibinfo {year}
  {2014})}\BibitemShut {NoStop}%
\bibitem [{\citenamefont {Rinc{\'o}n}\ \emph {et~al.}(2018)\citenamefont
  {Rinc{\'o}n}, \citenamefont {Dagotto},\ and\ \citenamefont
  {Feiguin}}]{Rincon:2018aa}%
  \BibitemOpen
  \bibfield  {author} {\bibinfo {author} {\bibfnamefont {J.}~\bibnamefont
  {Rinc{\'o}n}}, \bibinfo {author} {\bibfnamefont {E.}~\bibnamefont {Dagotto}},
  \ and\ \bibinfo {author} {\bibfnamefont {A.~E.}\ \bibnamefont {Feiguin}},\
  }\href {\doibase 10.1103/PhysRevB.97.235104} {\bibfield  {journal} {\bibinfo
  {journal} {Physical Review B}\ }\textbf {\bibinfo {volume} {97}},\ \bibinfo
  {pages} {235104} (\bibinfo {year} {2018})}\BibitemShut {NoStop}%
\bibitem [{\citenamefont {Eckstein}\ and\ \citenamefont
  {Kollar}(2008)}]{Eckstein:2008aa}%
  \BibitemOpen
  \bibfield  {author} {\bibinfo {author} {\bibfnamefont {M.}~\bibnamefont
  {Eckstein}}\ and\ \bibinfo {author} {\bibfnamefont {M.}~\bibnamefont
  {Kollar}},\ }\href {\doibase 10.1103/PhysRevB.78.205119} {\bibfield
  {journal} {\bibinfo  {journal} {Physical Review B}\ }\textbf {\bibinfo
  {volume} {78}},\ \bibinfo {pages} {205119} (\bibinfo {year}
  {2008})}\BibitemShut {NoStop}%
\bibitem [{\citenamefont {Shao}\ \emph {et~al.}(2016)\citenamefont {Shao},
  \citenamefont {Tohyama}, \citenamefont {Luo},\ and\ \citenamefont
  {Lu}}]{Shao:2016aa}%
  \BibitemOpen
  \bibfield  {author} {\bibinfo {author} {\bibfnamefont {C.}~\bibnamefont
  {Shao}}, \bibinfo {author} {\bibfnamefont {T.}~\bibnamefont {Tohyama}},
  \bibinfo {author} {\bibfnamefont {H.-G.}\ \bibnamefont {Luo}}, \ and\
  \bibinfo {author} {\bibfnamefont {H.}~\bibnamefont {Lu}},\ }\href {\doibase
  10.1103/PhysRevB.93.195144} {\bibfield  {journal} {\bibinfo  {journal}
  {Physical Review B}\ }\textbf {\bibinfo {volume} {93}},\ \bibinfo {pages}
  {195144} (\bibinfo {year} {2016})}\BibitemShut {NoStop}%
\bibitem [{\citenamefont {Lu}\ \emph {et~al.}(2015)\citenamefont {Lu},
  \citenamefont {Shao}, \citenamefont {Bon{\v c}a}, \citenamefont {Manske},\
  and\ \citenamefont {Tohyama}}]{Lu:2015aa}%
  \BibitemOpen
  \bibfield  {author} {\bibinfo {author} {\bibfnamefont {H.}~\bibnamefont
  {Lu}}, \bibinfo {author} {\bibfnamefont {C.}~\bibnamefont {Shao}}, \bibinfo
  {author} {\bibfnamefont {J.}~\bibnamefont {Bon{\v c}a}}, \bibinfo {author}
  {\bibfnamefont {D.}~\bibnamefont {Manske}}, \ and\ \bibinfo {author}
  {\bibfnamefont {T.}~\bibnamefont {Tohyama}},\ }\href {\doibase
  10.1103/PhysRevB.91.245117} {\bibfield  {journal} {\bibinfo  {journal}
  {Physical Review B}\ }\textbf {\bibinfo {volume} {91}},\ \bibinfo {pages}
  {245117} (\bibinfo {year} {2015})}\BibitemShut {NoStop}%
\bibitem [{\citenamefont {Paeckel}\ \emph {et~al.}(2020)\citenamefont
  {Paeckel}, \citenamefont {Fauseweh}, \citenamefont {Osterkorn}, \citenamefont
  {K{\"o}hler}, \citenamefont {Manske},\ and\ \citenamefont
  {Manmana}}]{Paeckel:2020aa}%
  \BibitemOpen
  \bibfield  {author} {\bibinfo {author} {\bibfnamefont {S.}~\bibnamefont
  {Paeckel}}, \bibinfo {author} {\bibfnamefont {B.}~\bibnamefont {Fauseweh}},
  \bibinfo {author} {\bibfnamefont {A.}~\bibnamefont {Osterkorn}}, \bibinfo
  {author} {\bibfnamefont {T.}~\bibnamefont {K{\"o}hler}}, \bibinfo {author}
  {\bibfnamefont {D.}~\bibnamefont {Manske}}, \ and\ \bibinfo {author}
  {\bibfnamefont {S.~R.}\ \bibnamefont {Manmana}},\ }\href {\doibase
  10.1103/PhysRevB.101.180507} {\bibfield  {journal} {\bibinfo  {journal}
  {Physical Review B}\ }\textbf {\bibinfo {volume} {101}},\ \bibinfo {pages}
  {180507(R)} (\bibinfo {year} {2020})}\BibitemShut {NoStop}%
\bibitem [{\citenamefont {Kogoj}\ \emph {et~al.}(2016)\citenamefont {Kogoj},
  \citenamefont {Vidmar}, \citenamefont {Mierzejewski}, \citenamefont
  {Trugman},\ and\ \citenamefont {Bon{\v c}a}}]{Kogoj:2016aa}%
  \BibitemOpen
  \bibfield  {author} {\bibinfo {author} {\bibfnamefont {J.}~\bibnamefont
  {Kogoj}}, \bibinfo {author} {\bibfnamefont {L.}~\bibnamefont {Vidmar}},
  \bibinfo {author} {\bibfnamefont {M.}~\bibnamefont {Mierzejewski}}, \bibinfo
  {author} {\bibfnamefont {S.~A.}\ \bibnamefont {Trugman}}, \ and\ \bibinfo
  {author} {\bibfnamefont {J.}~\bibnamefont {Bon{\v c}a}},\ }\href {\doibase
  10.1103/PhysRevB.94.014304} {\bibfield  {journal} {\bibinfo  {journal}
  {Physical Review B}\ }\textbf {\bibinfo {volume} {94}},\ \bibinfo {pages}
  {014304} (\bibinfo {year} {2016})}\BibitemShut {NoStop}%
\bibitem [{\citenamefont {White}(1992)}]{dmrg1}%
  \BibitemOpen
  \bibfield  {author} {\bibinfo {author} {\bibfnamefont {S.~R.}\ \bibnamefont
  {White}},\ }\href {\doibase 10.1103/PhysRevLett.69.2863} {\bibfield
  {journal} {\bibinfo  {journal} {Physical Review Letters}\ }\textbf {\bibinfo
  {volume} {69}},\ \bibinfo {pages} {2863} (\bibinfo {year}
  {1992})}\BibitemShut {NoStop}%
\bibitem [{\citenamefont {White}(1993)}]{dmrg2}%
  \BibitemOpen
  \bibfield  {author} {\bibinfo {author} {\bibfnamefont {S.~R.}\ \bibnamefont
  {White}},\ }\href {\doibase 10.1103/PhysRevB.48.10345} {\bibfield  {journal}
  {\bibinfo  {journal} {Physical Review B}\ }\textbf {\bibinfo {volume} {48}},\
  \bibinfo {pages} {10345} (\bibinfo {year} {1993})}\BibitemShut {NoStop}%
\bibitem [{\citenamefont {Schollw\"ock}(2005)}]{dmrg3}%
  \BibitemOpen
  \bibfield  {author} {\bibinfo {author} {\bibfnamefont {U.}~\bibnamefont
  {Schollw\"ock}},\ }\href {\doibase 10.1103/RevModPhys.77.259} {\bibfield
  {journal} {\bibinfo  {journal} {Reviews of Modern Physics}\ }\textbf
  {\bibinfo {volume} {77}},\ \bibinfo {pages} {259} (\bibinfo {year}
  {2005})}\BibitemShut {NoStop}%
\bibitem [{\citenamefont {Daley}\ \emph {et~al.}(2004)\citenamefont {Daley},
  \citenamefont {Kollath}, \citenamefont {Schollw{\"o}ck},\ and\ \citenamefont
  {Vidal}}]{Daley_2004}%
  \BibitemOpen
  \bibfield  {author} {\bibinfo {author} {\bibfnamefont {A.~J.}\ \bibnamefont
  {Daley}}, \bibinfo {author} {\bibfnamefont {C.}~\bibnamefont {Kollath}},
  \bibinfo {author} {\bibfnamefont {U.}~\bibnamefont {Schollw{\"o}ck}}, \ and\
  \bibinfo {author} {\bibfnamefont {G.}~\bibnamefont {Vidal}},\ }\href
  {\doibase 10.1088/1742-5468/2004/04/p04005} {\bibfield  {journal} {\bibinfo
  {journal} {Journal of Statistical Mechanics: Theory and Experiment}\ }\textbf
  {\bibinfo {volume} {2004}},\ \bibinfo {pages} {P04005} (\bibinfo {year}
  {2004})}\BibitemShut {NoStop}%
\bibitem [{\citenamefont {White}\ and\ \citenamefont
  {Feiguin}(2004)}]{White04}%
  \BibitemOpen
  \bibfield  {author} {\bibinfo {author} {\bibfnamefont {S.~R.}\ \bibnamefont
  {White}}\ and\ \bibinfo {author} {\bibfnamefont {A.~E.}\ \bibnamefont
  {Feiguin}},\ }\href {\doibase 10.1103/PhysRevLett.93.076401} {\bibfield
  {journal} {\bibinfo  {journal} {Physical Review Letters}\ }\textbf {\bibinfo
  {volume} {93}},\ \bibinfo {pages} {076401} (\bibinfo {year}
  {2004})}\BibitemShut {NoStop}%
\bibitem [{\citenamefont {Schollw{\"o}ck}(2011)}]{Uli_2011}%
  \BibitemOpen
  \bibfield  {author} {\bibinfo {author} {\bibfnamefont {U.}~\bibnamefont
  {Schollw{\"o}ck}},\ }\href {\doibase
  https://doi.org/10.1016/j.aop.2010.09.012} {\bibfield  {journal} {\bibinfo
  {journal} {Annals of Physics}\ }\textbf {\bibinfo {volume} {326}},\ \bibinfo
  {pages} {96 } (\bibinfo {year} {2011})}\BibitemShut {NoStop}%
\bibitem [{\citenamefont {Paeckel}\ \emph {et~al.}(2019)\citenamefont
  {Paeckel}, \citenamefont {K{\"o}hler}, \citenamefont {Swoboda}, \citenamefont
  {Manmana}, \citenamefont {Schollw{\"o}ck},\ and\ \citenamefont
  {Hubig}}]{Paeckel2019}%
  \BibitemOpen
  \bibfield  {author} {\bibinfo {author} {\bibfnamefont {S.}~\bibnamefont
  {Paeckel}}, \bibinfo {author} {\bibfnamefont {T.}~\bibnamefont {K{\"o}hler}},
  \bibinfo {author} {\bibfnamefont {A.}~\bibnamefont {Swoboda}}, \bibinfo
  {author} {\bibfnamefont {S.~R.}\ \bibnamefont {Manmana}}, \bibinfo {author}
  {\bibfnamefont {U.}~\bibnamefont {Schollw{\"o}ck}}, \ and\ \bibinfo {author}
  {\bibfnamefont {C.}~\bibnamefont {Hubig}},\ }\href {\doibase
  https://doi.org/10.1016/j.aop.2019.167998} {\bibfield  {journal} {\bibinfo
  {journal} {Annals of Physics}\ }\textbf {\bibinfo {volume} {411}},\ \bibinfo
  {pages} {167998} (\bibinfo {year} {2019})}\BibitemShut {NoStop}%
\bibitem [{\citenamefont {Jeckelmann}\ \emph {et~al.}(2000)\citenamefont
  {Jeckelmann}, \citenamefont {Gebhard},\ and\ \citenamefont
  {Essler}}]{Jeckelmann:2000aa}%
  \BibitemOpen
  \bibfield  {author} {\bibinfo {author} {\bibfnamefont {E.}~\bibnamefont
  {Jeckelmann}}, \bibinfo {author} {\bibfnamefont {F.}~\bibnamefont {Gebhard}},
  \ and\ \bibinfo {author} {\bibfnamefont {F.~H.~L.}\ \bibnamefont {Essler}},\
  }\href {\doibase 10.1103/PhysRevLett.85.3910} {\bibfield  {journal} {\bibinfo
   {journal} {Physical Review Letters}\ }\textbf {\bibinfo {volume} {85}},\
  \bibinfo {pages} {3910} (\bibinfo {year} {2000})}\BibitemShut {NoStop}%
\bibitem [{\citenamefont {Lenar{\v c}i{\v c}}\ \emph
  {et~al.}(2014)\citenamefont {Lenar{\v c}i{\v c}}, \citenamefont {Gole{\v z}},
  \citenamefont {Bon{\v c}a},\ and\ \citenamefont {Prelov{\v
  s}ek}}]{Lenarcic2014}%
  \BibitemOpen
  \bibfield  {author} {\bibinfo {author} {\bibfnamefont {Z.}~\bibnamefont
  {Lenar{\v c}i{\v c}}}, \bibinfo {author} {\bibfnamefont {D.}~\bibnamefont
  {Gole{\v z}}}, \bibinfo {author} {\bibfnamefont {J.}~\bibnamefont {Bon{\v
  c}a}}, \ and\ \bibinfo {author} {\bibfnamefont {P.}~\bibnamefont {Prelov{\v
  s}ek}},\ }\href {\doibase 10.1103/PhysRevB.89.125123} {\bibfield  {journal}
  {\bibinfo  {journal} {Physical Review B}\ }\textbf {\bibinfo {volume} {89}},\
  \bibinfo {pages} {125123} (\bibinfo {year} {2014})}\BibitemShut {NoStop}%
\bibitem [{\citenamefont {Eckstein}\ and\ \citenamefont
  {Werner}(2013)}]{Eckstein2013}%
  \BibitemOpen
  \bibfield  {author} {\bibinfo {author} {\bibfnamefont {M.}~\bibnamefont
  {Eckstein}}\ and\ \bibinfo {author} {\bibfnamefont {P.}~\bibnamefont
  {Werner}},\ }\href {\doibase 10.1103/PhysRevLett.110.126401} {\bibfield
  {journal} {\bibinfo  {journal} {Physical Review Letters}\ }\textbf {\bibinfo
  {volume} {110}},\ \bibinfo {pages} {126401} (\bibinfo {year}
  {2013})}\BibitemShut {NoStop}%
\bibitem [{\citenamefont {Rossini}\ \emph {et~al.}(2014)\citenamefont
  {Rossini}, \citenamefont {Fazio}, \citenamefont {Giovannetti},\ and\
  \citenamefont {Silva}}]{Rossini2014}%
  \BibitemOpen
  \bibfield  {author} {\bibinfo {author} {\bibfnamefont {D.}~\bibnamefont
  {Rossini}}, \bibinfo {author} {\bibfnamefont {R.}~\bibnamefont {Fazio}},
  \bibinfo {author} {\bibfnamefont {V.}~\bibnamefont {Giovannetti}}, \ and\
  \bibinfo {author} {\bibfnamefont {A.}~\bibnamefont {Silva}},\ }\href
  {\doibase 10.1209/0295-5075/107/30002} {\bibfield  {journal} {\bibinfo
  {journal} {{EPL} (Europhysics Letters)}\ }\textbf {\bibinfo {volume} {107}},\
  \bibinfo {pages} {30002} (\bibinfo {year} {2014})}\BibitemShut {NoStop}%
\bibitem [{\citenamefont {Ewing}(2016)}]{Ewing2016}%
  \BibitemOpen
  \bibfield  {author} {\bibinfo {author} {\bibfnamefont {G.~M.}\ \bibnamefont
  {Ewing}},\ }\href@noop {} {\emph {\bibinfo {title} {Calculus of Variations
  with Applications}}},\ \bibinfo {edition} {revised}\ ed.\ (\bibinfo
  {publisher} {Dover Publications},\ \bibinfo {address} {Mineola, New York},\
  \bibinfo {year} {2016})\BibitemShut {NoStop}%
\bibitem [{Note1()}]{Note1}%
  \BibitemOpen
  \bibinfo {note} {Note that if $J = J(\omega , E, E_\omega , t_{\protect \rm
  prb})$ is also a function of $E_\omega = \partial _\omega E$, variations of
  $E_\omega $ generalize~(\ref {eq:sigma}) to $$ \sigma (\omega , t_{\protect
  \rm prb}) = \protect \frac {\delta \protect \mathcal J}{\delta E} = \protect
  \frac {\partial J}{\partial E} - \protect \frac {d}{d\omega } \protect \frac
  {\partial J}{\partial E_\omega }. $$ \par The consequences of this extra term
  will be explored in future work.}\BibitemShut {Stop}%
\bibitem [{\citenamefont {Matsueda}\ \emph {et~al.}(2005)\citenamefont
  {Matsueda}, \citenamefont {Tohyama},\ and\ \citenamefont
  {Maekawa}}]{Matsueda:2005aa}%
  \BibitemOpen
  \bibfield  {author} {\bibinfo {author} {\bibfnamefont {H.}~\bibnamefont
  {Matsueda}}, \bibinfo {author} {\bibfnamefont {N.~B.~T.}\ \bibnamefont
  {Tohyama}}, \ and\ \bibinfo {author} {\bibfnamefont {S.}~\bibnamefont
  {Maekawa}},\ }\href {\doibase 10.1103/PhysRevB.72.075136} {\bibfield
  {journal} {\bibinfo  {journal} {Physical Review B}\ }\textbf {\bibinfo
  {volume} {72}},\ \bibinfo {pages} {075136} (\bibinfo {year}
  {2005})}\BibitemShut {NoStop}%
\bibitem [{\citenamefont {Kohno}(2010)}]{Kohno:2010aa}%
  \BibitemOpen
  \bibfield  {author} {\bibinfo {author} {\bibfnamefont {M.}~\bibnamefont
  {Kohno}},\ }\href {\doibase 10.1103/PhysRevLett.105.106402} {\bibfield
  {journal} {\bibinfo  {journal} {Physical Review Letters}\ }\textbf {\bibinfo
  {volume} {105}},\ \bibinfo {pages} {106402} (\bibinfo {year}
  {2010})}\BibitemShut {NoStop}%
\bibitem [{\citenamefont {Fano}(1961)}]{Fano:1961aa}%
  \BibitemOpen
  \bibfield  {author} {\bibinfo {author} {\bibfnamefont {U.}~\bibnamefont
  {Fano}},\ }\href {\doibase 10.1103/PhysRev.124.1866} {\bibfield  {journal}
  {\bibinfo  {journal} {Physical Review}\ }\textbf {\bibinfo {volume} {124}},\
  \bibinfo {pages} {1866} (\bibinfo {year} {1961})}\BibitemShut {NoStop}%
\bibitem [{Note2()}]{Note2}%
  \BibitemOpen
  \bibinfo {note} {Further study of this model is currently
  underway.}\BibitemShut {Stop}%
\end{thebibliography}%

\end{document}